%% file: main.tex
\documentclass[10pt,conference]{IEEEtran}
\IEEEoverridecommandlockouts

\usepackage{cite}
\usepackage{amsmath,amssymb,amsfonts}
\usepackage{algorithmic}
\usepackage[switch]{lineno} 

\usepackage{lipsum} 
\usepackage{graphicx}
\usepackage{textcomp}
\usepackage{xcolor}
\usepackage{caption,xspace}
\usepackage{pifont}
\usepackage{enumitem}
\usepackage{textcomp,balance}
\usepackage{bm}
\usepackage{amsfonts}
\usepackage{url}
\usepackage{multirow}
\usepackage{booktabs}
\usepackage{longtable}
\usepackage{graphicx}
\usepackage{subcaption}
\usepackage{subfloat}
\usepackage{numprint}
\usepackage{ragged2e}
\usepackage{setspace}
\usepackage{framed}
\usepackage{tcolorbox}
\usepackage{hyperref}
\usepackage[linesnumbered,ruled,vlined]{algorithm2e}

\usepackage{amsmath,amssymb}
\usepackage{algorithmic}
\usepackage{graphicx}
\usepackage{xcolor}
\usepackage{tikz}
\usepackage{listings,balance,lstautogobble}
\usepackage{cleveref}
\usepackage{orcidlink}
\usepackage{colortbl, adjustbox}

\Crefname{section}{Sec.}{Secs.}
\Crefname{figure}{Fig.}{Figs.}
\Crefname{equation}{Eq.}{Eq.}

\definecolor{ForestGreen}{RGB}{34,139,34}

\newcommand{\system}{\textsc{QTuner}\xspace}

\author{
\IEEEauthorblockN{
Ming Zhong\IEEEauthorrefmark{2}\,\orcidlink{0009-0002-7814-7523},
Xiangyu Ren\IEEEauthorrefmark{2}\, \orcidlink{0009-0006-8914-2188},
Jinglei Cheng\IEEEauthorrefmark{3}\, \orcidlink{0000-0001-9535-6672},
Shaohua Li\IEEEauthorrefmark{2}\, \orcidlink{0000-0001-7556-3615},
Zhiding Liang\IEEEauthorrefmark{2}\IEEEauthorrefmark{4}*\, \orcidlink{0000-0002-7568-0165}
\thanks{*: Corresponding Author.}
}
\IEEEauthorblockA{
\IEEEauthorrefmark{2}The Chinese University of Hong Kong, Hong Kong SAR, China,
\IEEEauthorrefmark{3}University of Pittsburgh, The United States
}
\IEEEauthorblockA{
\IEEEauthorrefmark{4}State Key Laboratory of Quantum Information Technologies and Materials (CUHK), Hong Kong SAR, China\\
mzhong25@cse.cuhk.edu.hk, xiyu.ren@outlook.com, chengjl14@outlook.com, \\
shaohuali@cse.cuhk.edu.hk, zliang@cse.cuhk.edu.hk
}
}

\def\BibTeX{{\rm B\kern-.05em{\sc i\kern-.025em b}\kern-.08em
    T\kern-.1667em\lower.7ex\hbox{E}\kern-.125emX}}
    
\begin{document}

\title{\system: A Feature-Guided Adaptive Framework for Quantum-Compiler Optimization Pass Tuning}

\maketitle

\thispagestyle{plain}
\pagestyle{plain}

\begin{abstract}

Quantum compiler optimization is important for reducing cost and improving the execution fidelity of quantum circuits. In practice, this optimization is often formulated as \emph{reducing} circuit-level metrics. In classical compilers, pass tuning has been well-established through various methods. However, these methods cannot be directly applied to quantum compilers, because quantum programs define computations through quantum circuits and exhibit quantum-specific features. Existing quantum compiler tuning studies have attempted to incorporate quantum circuit features with compiler tuning techniques, but they still suffer from two key limitations. First, they tune only a limited subset of optimization passes, leaving the global effectiveness of the full optimization pass space underexplored. Second, the quantum features they use are mainly static features, such as structure and DAG features, which are not directly tied to the optimization characteristics of quantum circuits.

In this paper, we propose \system, a feature-guided framework for quantum compiler pass tuning that explores the full optimization pass space and can be adapted to different compilers and tuning objectives. We first construct an optimization dataset by performing Bayesian Optimization on 8,111 circuits and recording their optimized pass sequences. Based on this dataset, \system represents each circuit with two types of features: static circuit features that describe circuit structure and optimization-aware pass embeddings that capture circuit-specific optimization behavior. Finally, 
\system trains two models offline, which are used to efficiently retrieve and rank promising pass sequences for a new circuit. We evaluate \system on two benchmarks across Qiskit and PyTKET compilers.  On Qiskit, \system achieves up to a 10.6-percentage-point improvement in two-qubit (2Q) gate reduction over the strongest baseline while reducing tuning time by 73.59\%. On PyTKET, \system achieves up to a 9.4-percentage-point improvement in 2Q gate reduction and reduces tuning time by 64.49\%. These results demonstrate that \system enables efficient pass tuning across quantum compilers.

\end{abstract}


\input{ch1-intro}

\input{ch2-motivation}
\input{ch3-method}

\input{ch4-implementations}

\input{ch5-evaluation}

\input{ch6-related_work}
\input{ch7-conclusion}

\clearpage
\bibliographystyle{plain}
\bibliography{bibfile}

\end{document}

%% file: ch1-intro.tex
\section{Introduction}\label{sec:intro}

Quantum computing has emerged as a promising paradigm for solving problems intractable to classical computers~\cite{PhysRevLett_qa,  quantum_finance,liang2024combining}. Similar to software development for classical computers, quantum programming relies on software development kits (SDKs) and compilers, such as Qiskit~\cite{javadiabhari2024quantumcomputingqiskit} and PyTKET~\cite{Sivarajah_2021}, to implement quantum algorithms and execute them on quantum computers. In classical computing, compilers optimize programs by reducing code size and execution time~\cite{GAForReduceCodeSize,boca,PDCAT}, typically guided by program features such as control flow, data dependencies, and loop structures.

\begin{figure}[t]
    \centering
    \begin{subfigure}[t]{\linewidth}
        \centering
        \includegraphics[width=\linewidth]{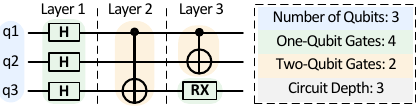}
        \caption{An example quantum circuit with quantum-specific features.}
        \label{fig:example_circuit}
    \end{subfigure}
    \\
    \begin{subfigure}[t]{\linewidth}
        \centering
        \includegraphics[width=\linewidth]{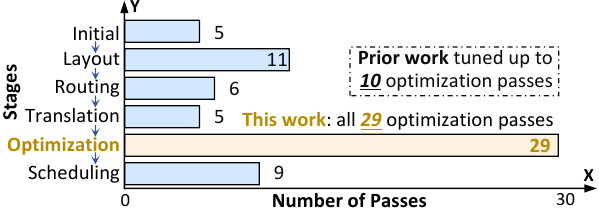}
        \caption{Compilation stages of the Qiskit compiler (V2.2.3).}
        \label{fig:compiler_stage}
    \end{subfigure}
    \caption{Challenges of quantum compiler pass tuning. (a) Quantum circuits have specific features distinct from classical programs, such as gate counts and circuit depth. (b) Existing studies do not cover the full optimization pass tuning space. }
    \label{fig:compiler_tuning_sdk}
    \vspace{-2.ex}
\end{figure}

However, quantum programs typically define computations in the form of quantum circuits, whose characteristics are fundamentally different from those of classical programs. As shown in \Cref{fig:example_circuit}, a quantum circuit consists of qubits and quantum gates, and its computation is realized through a sequence of one-qubit (1Q) and two-qubit (2Q) quantum operations. Therefore, quantum compilation transforms circuits into hardware executable forms through qubit mapping, routing, scheduling, and circuit optimization. Instead of focusing mainly on classical execution time, quantum compilers aim to \textbf{\emph{reduce}} circuit-level metrics such as 2Q gate count, 1Q gate count, and circuit depth. Reducing these metrics can mitigate accumulated gate errors and decoherence, thereby improving the execution fidelity of quantum programs~\cite{guoq,SABRE}.

As illustrated in \Cref{fig:compiler_stage}, a quantum compiler (e.g., Qiskit) consists of stages including initial, layout, routing, translation, optimization, and scheduling. Initial stage lowers multi-qubit operations down to a series of 1Q and 2Q operations. Layout maps logical qubits in quantum circuits to physical qubits in quantum backends, routing enables interactions between unconnected qubits, translation converts gates to hardware-supported bases, optimization applies analysis and transformation passes to reduce circuit cost, and scheduling assigns execution times to operations under hardware constraints. Among these stages, the optimization stage contains the largest number of passes (e.g., 29 in \Cref{fig:compiler_stage}), making it the most complex component. In contrast, other stages such as layout and routing typically rely on well-established heuristics, such as SABRE~\cite{SABRE}, which balances performance and efficiency.

Additionally, both classical and quantum compilers typically provide a set of predefined optimization levels, such as $O_3$, to enable users to quickly apply compilation optimizations. Each optimization level corresponds to a fixed pass sequence designed for general use. However, such fixed pass sequences may not achieve optimal performance for every input, because different inputs can exhibit diverse structural characteristics and optimization opportunities. Therefore, compiler pass tuning aims to identify effective pass sequences, including both pass combinations and orderings, for each input program or circuit, thereby improving the compilation performance.

In classical compilers, prior studies have investigated both compiler pass and option tuning. Existing approaches include heuristic search methods such as Bayesian optimization (BO) and genetic algorithms (GA)~\cite{boca,cobyan,PDCAT,gene1,gene2}, reinforcement learning (RL) for sequential pass selection~\cite{rl1,rl3,rl4}, and similarity-based retrieval using program features~\cite{ICS_Similarity,SC_similarity}. Recent studies further leverage large language models (LLMs) combined with retrieval-augmented-generation (RAG) techniques to guide compiler optimization through their reasoning capabilities~\cite{pan2026eccoevidencedrivencausalreasoning,qiu2026passbypassoptimizationintentdrivenir,meta_llm_compiler}. Some studies use the term ``phase-ordering" to emphasize the ordering aspect of pass tuning~\cite{cc_phase,oopsla_phase,autophase,ipdps_phase,internetware_phase}.

In quantum compiler pass tuning, existing studies have attempted to adapt heuristic  and RL methods from traditional compiler tuning to quantum circuits by using partial circuit-level features, like structure or DAG features ~\cite{DAC23,quantum_tuning_cibda,mills2026reinforcementlearningadaptivecomposition,dangwal2025cliffordassistedoptimalpass,mqtpredictor,tuniq}. However, their optimization pass coverage remains limited. For example, \cite{mills2026reinforcementlearningadaptivecomposition} explores only four optimization passes in PyTKET V2.9.3. Although \cite{DAC23} and \cite{tuniq} tune passes across multiple compilation stages, their optimization-stage search spaces cover only 9 out of 23 optimization passes in Qiskit V0.39.2 and 10 out of 28 optimization passes in Qiskit V2.1.2, respectively. These studies therefore do not fully cover the quantum compiler optimization-pass space.

In this paper, we present the first tuning work that covers the full optimization pass space of studied quantum compilers. We propose a feature-guided framework, namely \system, that extracts both structural and optimization-aware features from quantum circuits. Based on these features, we train graph transformer-based models to learn circuit representations, retrieve similar circuits from a large-scale optimization dataset, and identify promising optimization pass sequences. We further apply lightweight BO to refine the retrieved sequences, achieving better optimization performance while balancing the time cost of the optimization pass tuning process.

Specifically, we first construct an optimization dataset by performing BO tuning over all optimization passes in Qiskit V2.2.3 (the latest version available when this work began) on 8,111 circuits from QCircuitBench~\cite{yang2025qcircuitbench}. Based on this dataset, \system represents each circuit's optimization characteristics by combining static circuit features with optimization-aware pass embeddings obtained through pass profiling. For a new input circuit, \system trains a pruning model and a ranker model to retrieve and rank promising pass sequences from the dataset, followed by lightweight BO refinement to further explore the optimization space beyond the optimization dataset and produce the final optimized pass sequence.

Our four contributions are summarized as follows:

\begin{itemize}

    \item We construct a large-scale optimization dataset by performing BO-tuning on the QCircuitBench quantum circuit dataset~\cite{yang2025qcircuitbench}, providing a promising foundation for quantum compiler optimization  pass tuning.

    \item We introduce optimization-aware pass embeddings through the pass profiling mechanism, and combine pass embeddings with static circuit features to model circuit-specific optimization characteristics.
    
    \item We propose \system, a feature-guided framework that retrieves, ranks, and refines optimization pass sequences, achieving strong optimization effectiveness and time efficiency across quantum compilers and tuning objectives.
    
    \item We evaluate \system on two benchmarks comprising 180 cases. Compared with the strongest baseline, \system achieves up to a 10.6-percentage-point improvement in 2Q gate reduction while reducing tuning time by 73.59\% on Qiskit. On PyTKET, \system achieves up to a 9.4-percentage-point improvement in 2Q gate reduction and reduces tuning time by 64.49\%.
    
\end{itemize}

%% file: ch2-motivation.tex
\section{Motivation}\label{Sec:mot}

Compiler pass tuning is well established in classical compilers and is also important for quantum compilers. However, as discussed in \Cref{sec:intro}, existing quantum compiler pass tuning studies remain limited due to insufficient optimization-pass coverage and partial circuit feature modeling. Therefore, we detail our motivation from two aspects: exploring the full optimization pass space and modeling circuit features more comprehensively to better capture optimization behavior.

\subsection{Exploring the Full Optimization Pass Space}\label{sec:mot_opt}

As introduced in \Cref{sec:intro}, existing studies explore only a limited subset of optimization passes in quantum compilers. This is mainly because they usually target specific optimization metrics, such as 2Q gate reduction, and therefore select only passes directly related to these objectives. For example, the work~\cite{mills2026reinforcementlearningadaptivecomposition} focuses on reducing 2Q gates and thus tunes only four 2Q-related optimization passes in PyTKET. Other studies also tune layout and routing options together with optimization passes~\cite{DAC23,mqtpredictor,tuniq}. In such cases, covering all optimization passes would further enlarge the search space, so these methods restrict tuning to a selected subset of passes.

However, tuning only passes directly related to a single metric, or only a subset of optimization passes, can overlook global effectiveness. Since the effectiveness of a pass often depends on its cooperation with other passes and its position in the global pass sequence, exploring the full optimization pass space is crucial for discovering globally effective sequences and further improving quantum circuit optimization.

\subsection{Comprehensive Feature Modeling of Quantum Circuits}

In classical compilers, a program can be characterized by various features, such as control flow and data dependencies. However, these features cannot be directly applied to quantum circuits due to the fundamental differences in quantum computation. Existing quantum compiler pass tuning studies~\cite{mqtpredictor,DAC23,quantum_tuning_cibda,mills2026reinforcementlearningadaptivecomposition,tuniq} have therefore adopted specific types of quantum circuit features, such as structure features including entanglement ratio, liveness, and parallelism, or DAG features that model qubit-gate dependencies as a directed acyclic graph. Although these features can effectively characterize the static properties of quantum circuits, they cannot directly capture how a circuit responds to different optimization passes.

To address this limitation, we are motivated to directly measure how a quantum circuit responds to different passes, leading to a pass profiling mechanism. Specifically, we independently apply each optimization pass to a circuit and calculate the changes in circuit metrics, before and after applying the pass. These changes reflect the optimization effect of each pass on the circuit. We then concatenate the metric changes across all passes to obtain the pass embedding for each circuit. Compared with static circuit features, pass embeddings provide more direct optimization-aware features.

Although quantum compiler optimization passes can interact with each other, profiling pass combinations would substantially increase both the profiling cost and the embedding dimension. Therefore, we profile each pass independently as a practical trade-off between capturing optimization-aware features and keeping the profiling process efficient and scalable.

%% file: ch3-method.tex
\section{Our Methodology: \system}\label{sec:method}

\begin{figure}[t]
\includegraphics[width=\linewidth]{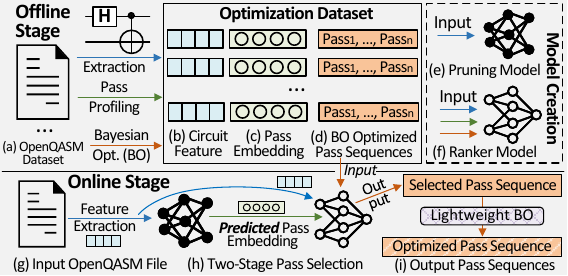}
\caption{The overall workflow of \system. In the figure, $\textcolor{blue}{\rightarrow}$ denotes the flow of circuit features, $\textcolor{green!50!black}{\rightarrow}$ denotes pass embeddings, and $\textcolor{orange}{\rightarrow}$ denotes optimized pass sequences.}
\label{fig:workflow}
\vspace{-2.ex}
\end{figure}

\subsection{Overview of \system}\label{sec:overview}

\Cref{fig:workflow} illustrates the overall workflow of \system, which consists of an offline stage and an online stage. In the offline stage, \system constructs an optimization dataset containing various quantum circuits. For each circuit, it extracts static circuit features, profiles pass embeddings, and records BO-optimized pass sequences (\Cref{fig:workflow}(b)--(d)). Based on this dataset, \system trains two models for pass sequence selection: a pruning model for efficient candidate retrieval and a ranker model for candidate sequence ranking (\Cref{fig:workflow}(e)--(f)).

In the online stage, \system takes a quantum circuit as input and first extracts its static circuit features. The pruning model then predicts its pass embedding and retrieves a small set of similar circuits from the optimization dataset. The optimized pass sequences of these retrieved circuits are used as candidates and scored by the ranker model. Finally, the highest-ranked valid sequence is refined by lightweight BO to generate the final optimized pass sequence (\Cref{fig:workflow}(h)--(i)).

\subsection{Offline Stage}\label{sec:offline}

\subsubsection{Tuning Objective}\label{sec:tuning_obj}
Existing quantum compiler pass tuning works consider various evaluation metrics including the 2Q gate counts~\cite{quantum_tuning_cibda,mills2026reinforcementlearningadaptivecomposition}, 1Q gate counts~\cite{quantum_tuning_cibda}, circuit depth~\cite{quantum_tuning_cibda,mqtpredictor,DAC23}, and estimated fidelity~\cite{DAC23,mqtpredictor,tuniq}.

In this work, we adopt a weighted tuning objective that jointly considers 2Q gate count, 1Q gate count, and circuit depth:
$Objective = \alpha \cdot \Delta_{\text{2Q}} + \beta \cdot \Delta_{\text{1Q}} + \gamma \cdot \Delta_{\text{Depth}}$, where $\Delta_{\text{2Q}}$, $\Delta_{\text{1Q}}$, and $\Delta_{\text{Depth}}$ 
denote the \emph{\textbf{reductions}} in 2Q gate count, 1Q gate count, and circuit depth over $O_3$, respectively.

This comprehensive objective provides a straightforward evaluation of circuit optimization quality across circuit-level metrics. We do not use estimated fidelity~\cite{DAC23,mqtpredictor,tuniq} as the tuning objective, which estimates fidelity by multiplying the backend-specific fidelity of each gate in the compiled circuit. Since it is tightly coupled with a particular quantum device, we instead optimize hardware-agnostic circuit-level metrics, which are directly affected by optimization-stage pass tuning. Notably, we further evaluate \system under an estimated-fidelity tuning objective in \Cref{sec:efp}, demonstrating that \system can support different tuning objectives.

\begin{figure}[t]
\includegraphics[width=\linewidth]{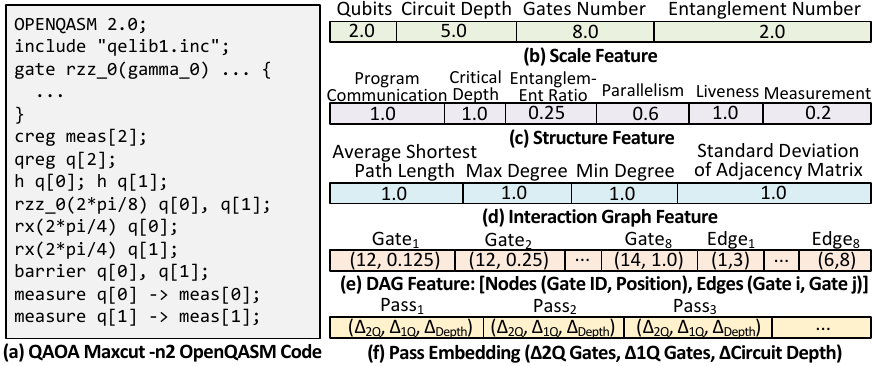}
\caption{Feature representations of an example quantum circuit, QAOA MaxCut-n2, in OpenQASM. OpenQASM is a unified representation for quantum circuits~\cite{openqasm2}.}
\label{fig:feature_representation}
\vspace{-2.ex}
\end{figure}

\subsubsection{Static Circuit Feature Representation}\label{sec:features}

Quantum circuit features are essential for modeling optimization behavior. Unlike classical programs, quantum circuits run on noisy and resource-constrained quantum devices, where hardware constraints such as limited qubit connectivity can significantly affect compilation results~\cite{supermarq,inter_graph}. Thus, different circuits may exhibit different optimization behaviors even under the same compiler and backend, making effective feature modeling necessary for selecting suitable optimization strategies.

To effectively optimize a quantum circuit, we characterize it from multiple perspectives, including circuit scale, operation structure, qubit interactions, and gate dependencies. Circuit scale reflects the basic optimization difficulty, while operation structure captures properties such as entanglement, parallelism, and qubit liveness. Qubit interactions, mainly induced by 2Q gates, are critical because 2Q gates are typically more costly and error-prone than 1Q gates. Gate dependencies further determine execution order and potential simplification opportunities. Together, these features provide a comprehensive circuit representation for modeling optimization similarity.

Inspired by prior works on characterizing quantum circuits~\cite{supermarq,inter_graph,quest,tosem_circuits}, we unify four types of static circuit features. We use a simplified quantum circuit (\Cref{fig:feature_representation}(a)) as an example to illustrate the four types of  selected features.

The four types of static circuit features and their computation methods are detailed as follows:

\begin{enumerate}[leftmargin=*]
    \item \textbf{Scale features}. Scale features describe the basic size of a quantum circuit, including the number of qubits, circuit depth, total gate count, and the number of entangling operations, as shown in \Cref{fig:feature_representation}(b). These features are directly extracted from the circuit by counting the corresponding circuit elements and the circuit depth.

    \item \textbf{Structure features}. Structure features describe the organization and resource usage patterns of quantum operations. Following SupermarQ~\cite{supermarq}, we use all six structure features, including program communication, critical depth, entanglement ratio, parallelism, liveness, and measurement, as shown in \Cref{fig:feature_representation}(c). These features are computed by analyzing qubit interactions, gate dependencies, 2Q gates ratio, operation parallelism, qubit activity, and measurement usage. Specifically, we follow the feature definitions and computation formulas proposed in SupermarQ~\cite{supermarq}.

     \item \textbf{Interaction graph features}. Interaction graph features capture the interaction patterns among qubits, including average shortest path length, maximum degree, minimum degree, and the standard deviation of the adjacency matrix, as shown in \Cref{fig:feature_representation}(d). We first convert each quantum circuit into a qubit interaction graph, where each node represents a qubit and each edge represents a 2Q operation between two qubits. If multiple 2Q operations are applied to the same qubit pair, the corresponding edge weight records the number of interactions. Based on this interaction graph, we compute the four interaction graph features. These four features are selected following the empirical analysis in~\cite{inter_graph}, which shows that they are informative enough for characterizing quantum circuit interaction behaviors.
    
    \item \textbf{DAG features}. DAG features capture the execution dependency relationships among quantum gates in a quantum circuit~\cite{tosem_circuits,quest}. We convert each quantum circuit into a circuit DAG, where each node represents a quantum gate and each directed edge represents a dependency between two gates. Specifically, an edge $(g_i, g_j)$ indicates that gate $g_i$ must be executed before gate $g_j$, because they operate on at least one common qubit and follow this order in the original circuit. Each node is encoded with a gate-type index (gate ID) and a normalized positional value, as shown in \Cref{fig:feature_representation}(e), where the gate-type index is later mapped to a learnable embedding in the model and the positional value indicates the relative order of the gate in the circuit DAG. For the edge structure, we use an edge list of gate-index pairs to represent gate dependencies.
    
\end{enumerate}

These four types of static circuit features jointly capture various circuit characteristics, but they do not explicitly reflect how circuits respond to optimization passes, which motivates our design of pass embeddings introduced in \Cref{sec:pass_embeddings}.

\begin{figure}[t]
\includegraphics[width=\linewidth]{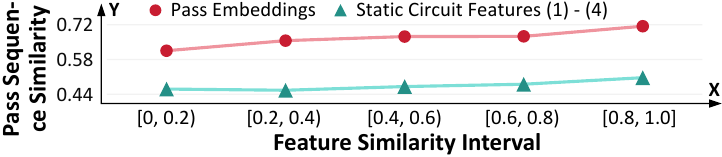}
\caption{Static feature-based and pass embedding-based measurements of BO optimized pass sequence similarity (measured by edit similarity, higher is better).}
\label{fig:cmp_pass_attr}
\vspace{-2.ex}
\end{figure}

\subsubsection{Optimization Dataset}\label{sec:dataset}

The optimization dataset is built from QCircuitBench~\cite{yang2025qcircuitbench}, a large-scale quantum circuit dataset, and serves as the retrieval basis for \system. Each circuit is represented by static circuit features (\Cref{sec:features}) and pass embeddings (\Cref{sec:pass_embeddings}), associated with its best BO-optimized pass sequence. These BO-optimized pass sequences are obtained through long-term BO tuning on the QCircuitBench in the offline stage. In the online stage, \system retrieves similar circuits from this dataset and uses their optimized pass sequences as candidates for refinement. The detailed dataset construction process is described in \Cref{sec:imp_dataset}.

\subsubsection{Pass Embeddings}\label{sec:pass_embeddings}

To better model the optimization behavior of quantum circuits beyond static features, we introduce pass embeddings, as shown in \Cref{fig:feature_representation}(f). Specifically, we apply each optimization pass individually to a circuit and record the relative changes in three metrics: 2Q gate count, 1Q gate count, and circuit depth, e.g., ($\Delta 2Q$, $\Delta 1Q$, $\Delta Depth$). The changes across all passes are concatenated to form a pass embedding.

It is also worth noting that pass profiling is time-consuming. For each circuit, it requires independently applying each optimization pass once to obtain its pass embedding. Therefore, pass profiling is \emph{\textbf{only}} performed in the offline stage to obtain accurate pass embeddings for circuits in the optimization dataset. In the online stage, \system uses a trained model (the pruning model) to predict the pass embedding of a new circuit from its static features, avoiding the high computational overhead of applying all optimization passes individually.

We next empirically demonstrate that pass embeddings provide a more effective representation of the relationship between circuits and optimization passes, compared to four types of static circuit features.
Specifically, we aim to examine whether pass embeddings or static circuit features better reflect the similarity of optimized pass sequences.
Our intuition is that a useful representation for retrieval should assign higher similarity to circuit pairs whose optimized pass sequences are also similar. To evaluate this, we first randomly sample 2,000 circuits from QCircuitBench and construct all circuit pairs among them, resulting in 4,000,000 pairs in total. We use sampling because constructing pairs from all circuits would produce more than 64 million pairs, leading to prohibitively high analysis cost. For each pair, we compute two inner-product similarity scores (normalized to [0,1] using min-max): one between their pass embeddings and the other between their static circuit features, where static circuit features are obtained by concatenating circuit features (1)--(4). We then group these similarity scores into intervals with a step size of 0.2 and compute, for each interval, the average edit similarity~\cite{EditDis} between the corresponding optimized pass sequences.

In \Cref{fig:cmp_pass_attr}, we observe that as the similarity between pass embeddings increases, the edit similarity between the corresponding pass sequences also consistently increases. Moreover, across all intervals, the pass sequence similarity corresponding to the pass embedding similarity is consistently higher than that corresponding to static circuit feature similarity. These results suggest that pass embeddings are more correlated with pass sequence similarity than static circuit features.

\begin{figure*}[t]
\centering
\includegraphics[width=\linewidth]{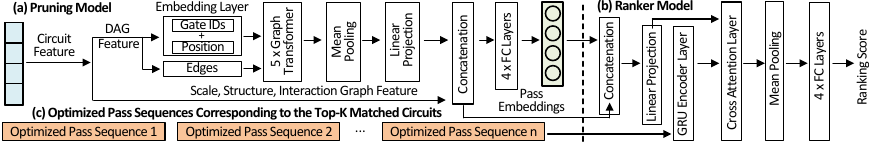}
\caption{Two-stage selection model design of \system, including the pruning model (a) and the ranker model (b).}
\label{fig:model_architecture}
\vspace{-3.ex}
\end{figure*}

\subsubsection{Model Creation}\label{sec:model}

For a new circuit, \system needs to match it with circuits in the optimization dataset and retrieve their corresponding optimized pass sequences as candidates. A straightforward solution is to train a model to score all optimized pass sequences in the dataset and select the best one. However, ranking all pass sequences for every input would introduce substantial computational and time overhead. Therefore, \system adopts a two-stage selection strategy. The pruning model first retrieves a small set of similar circuits from the optimization dataset, and the ranker model then scores the optimized pass sequences associated with these circuits.

Specifically, as illustrated in \Cref{fig:model_architecture}, the pruning model predicts the pass embedding of the new circuit, thereby avoiding the costly process of pass profiling. Based on the predicted pass embedding and the profiled pass embeddings of circuits in the optimization dataset, \system computes their inner-product similarities. It then selects the top-$k$ most similar circuits according to the similarity scores and retrieves their corresponding optimized pass sequences from the dataset.

Next, the ranker model takes the static circuit features, the predicted pass embedding, and the retrieved top-$k$ candidate pass sequences as input. It scores these candidates according to their \emph{relative} optimization effectiveness for the input circuit, i.e., their ranking among the $k$ candidate pass sequences, and selects the most promising one for further refinement.

We describe the two models in detail as follows.

\begin{itemize}[leftmargin=*]
    \item \textbf{Pruning Model}. 
    As shown in \Cref{fig:model_architecture}(a), the pruning model takes circuit features as input and predicts the pass embeddings. It first encodes the gate ID and normalized position of each node in the DAG features into node embeddings, and then feeds the node embeddings together with the DAG edge list into graph transformer layers. The graph transformer produces DAG representations, which are then aggregated by mean pooling and projected into a global DAG-level representation. This representation is then concatenated with scale, structure, and interaction graph features to form a unified circuit embedding. Finally, four fully connected (FC) layers are used to predict the pass embedding. This graph-transformer-based design has been adopted in existing studies to capture DAG-level circuit features~\cite{tosem_circuits,quest}.

    \item \textbf{Ranker Model}. 
    As shown in \Cref{fig:model_architecture}(b), the ranker model ranks the relative optimization effectiveness of pass sequences from the top-$k$ candidates (\Cref{fig:model_architecture}(c)). It takes static circuit features, predicted pass embeddings, and top-$k$ candidate pass sequences as input. The circuit features and predicted pass embeddings are concatenated and projected into a unified circuit-level representation, while each candidate sequence is encoded by a GRU encoder to capture pass ordering and composition. The circuit-level representation then interacts with each candidate sequence representation through a cross-attention layer to estimate the candidate's suitability for the input circuit. The attended representations are aggregated by mean pooling and passed through four FC layers to produce one ranking score for each candidate sequence. Since these ranking scores are used to compare the relative effectiveness of the top-$k$ candidates rather than predict absolute optimization scores, we train the ranker as a $k$-way candidate selection task with cross-entropy loss.
\end{itemize}

The trained two-stage models can be directly deployed in the online stage for effective pass retrieval for new circuits.

\subsection{Online Stage}
In the online stage, \system takes a quantum circuit as input (\Cref{fig:workflow}(g)) and extracts its static circuit features. The pruning model predicts its pass embedding and retrieves the top-$k$ candidate pass sequences from the optimization dataset. The ranker model then scores these candidates, and \system selects the highest-ranked valid sequence as the seed for lightweight BO refinement (\Cref{fig:workflow}(h)). If no valid candidate is found, \system falls back to the default $O_3$ sequence. The lightweight BO further explores the input-circuit-specific optimization space, rather than being limited to the optimized pass sequences retrieved from the dataset (\Cref{fig:workflow}(i)).

%% file: ch4-implementations.tex
\section{Implementations}\label{sec:imple}

\subsection{Offline Stage}
\subsubsection{SDK and Optimization Pass Selection}
We select Qiskit compiler V2.2.3 as the base compiler for tuning. The optimization stage of Qiskit V2.2.3 contains 29 passes, as listed in the official documentation~\cite{Web:qiskit_opt_pass}. We carefully examine the documentation to ensure that the selected passes do not alter circuit semantics. Qiskit explicitly marks passes that may change semantics, such as \texttt{RemoveBarriers}~\cite{Web:removebarriers}, which is not included in the optimization stage. Since none of the 29 optimization passes are marked as semantics-changing, we use all of them as the optimization pass tuning search space. 

Notably, the \texttt{ElidePermutations} pass in the optimization stage must run before the layout stage, as it becomes a no-op afterward due to physical connectivity constraints, according to the official documentation~\cite{Web:qiskit_opt_pass}. We retain it in the search space, although it has no effect in our setting. Additionally, due to their high algorithm complexity, we disable \texttt{HoareOptimizer} and \texttt{TemplateOptimization} for circuits with more than 1,000 initial gates.

\subsubsection{Simulator and Compilation Settings}
We use the FakeWashingtonV2 backend as our primary simulator. It models a 127-qubit heavy-hex device and has been widely adopted as an evaluation simulator in related studies~\cite{tosem_circuits,quest}.
Across all experiments, including optimization dataset construction, BO tuning, and evaluation, the layout, routing, translation, and scheduling configurations in the quantum compiler follow the default highest optimization levels (e.g., $O_3$ in Qiskit).

\subsubsection{Tuning Objective}

In \Cref{sec:tuning_obj}, we adopt a composite metric based on the reductions in 2Q gates, 1Q gates, and circuit depth. Since prior studies have shown that 2Q gates are more costly and error-prone, and thus dominate execution cost~\cite{guoq}, we assign weights of 80\%, 10\%, and 10\% to the three metrics, respectively, as shown in \Cref{eq:weighted_reduction}, where $S(c)$ calculates the weighted average reduction of the compiled circuit $c$ produced by the optimized pass sequence over the  $O_3$ baseline, and $N$ represents the number of 2Q or 1Q gates.

{\scriptsize
\setlength{\abovedisplayskip}{3pt}
\setlength{\belowdisplayskip}{3pt}
\setlength{\abovedisplayshortskip}{2pt}
\setlength{\belowdisplayshortskip}{2pt}
\begin{equation}
\label{eq:weighted_reduction}
\begin{aligned}
\mathrm{S}(c) ={}&
80\% \cdot \left(1-\frac{N_{2Q}(c)}{\max(N_{2Q}^{\mathrm{O_3}},1)}\right)
&+
10\% \cdot \left(1-\frac{N_{1Q}(c)}{\max(N_{1Q}^{\mathrm{O_3}},1)}\right) \\
&+
10\% \cdot \left(1-\frac{\mathrm{Depth}(c)}{\max(\mathrm{Depth}^{\mathrm{O_3}},1)}\right)
\end{aligned}
\end{equation}
}

\subsubsection{Pass Embeddings}
To profile the pass embeddings, we first compile each circuit with \texttt{optimization\_level=0}. We then apply each optimization pass individually to the compiled circuit and compute its relative metric changes as $\Delta m_p=(m_{\mathrm{before}}-m_{\mathrm{after}})/\max(m_{\mathrm{before}},1)$, where $m$ denotes 2Q gate count, 1Q gate count, or circuit depth, and $m_{\mathrm{before}}$ and $m_{\mathrm{after}}$ are measured before and after applying pass $p$, respectively. If a pass fails, its embedding entry is set to $(0.0,0.0,0.0)$. The pass embedding is formed by concatenating the three relative changes of all optimization passes.

\subsubsection{Optimization Dataset}\label{sec:imp_dataset}
We choose the QCircuitBench~\cite{yang2025qcircuitbench} as our dataset. QCircuitBench includes implementations of representative quantum algorithms as well as a large collection of randomly generated circuits, provided in OpenQASM formats (both in OpenQASM 2.0 and 3.0). OpenQASM serves as a unified representation and a common input format for different quantum SDKs and compilers~\cite{openqasm2}.

We first convert all OpenQASM 3.0 circuits to OpenQASM 2.0 using the \texttt{open\_qasm\_file\_conversion\_3\_to\_2} API from the ``mpqp" library in Python 3.11, as OpenQASM 2.0 is more compatible across different SDKs~\cite{bench_qasm}. To match the backend constraints, we retain only circuits with no more than 127 qubits, resulting in a final dataset of 8,111 circuits.
For each circuit, we extract the four types of circuit features in \Cref{sec:features} and obtain pass embeddings via pass profiling.

Next, we implement BO as a configuration-space search over optimization pass sequences. Each sequence is encoded by the selected passes, their relative ordering, and their repetition counts. The final sequence is obtained by expanding selected passes according to their repetition counts and sorting them by priority. We allow each pass to appear up to three times, as repeated applications may expose additional optimization opportunities. During tuning, BO starts from the default $O_3$ sequence and 19 random configurations, resulting in 20 initial samples in total. It then iteratively fits a Gaussian-process surrogate model and selects candidates by maximizing the expected-improvement acquisition function. Each candidate is evaluated using \Cref{eq:weighted_reduction}. For the offline dataset construction stage, we run BO for 500 iterations with early stopping after 50 consecutive rounds without improvement.

Following prior work~\cite{quantum_tuning_cibda,mqtpredictor}, we further check the hardware validity of each candidate pass sequence, including backend gate-set compatibility and 2Q topology constraints of the compiled circuit.If a candidate fails to satisfy any of these hardware constraints, we consider the corresponding pass sequence unsuccessful. Finally, we obtain 8,111 optimized pass sequences as the foundation optimization dataset.

Constructing the optimization dataset requires 447 CPU-core hours of BO tuning in total, measured as accumulated time over all circuits. Since per-circuit tuning tasks are independent, we parallelize them using 10 workers, resulting in approximately 72 hours of wall-clock time due to workload imbalance, as some circuits require long tuning time.

\subsubsection{Model Creation}
In \Cref{fig:model_architecture}(a), the pruning model first maps DAG features to 192-dimensional embeddings and processes them with five graph transformer layers. The resulting representations are mean-pooled and projected to 256 dimensions, then concatenated with the other three circuit features to form a 270-dimensional circuit representation. Finally, four FC layers map this representation to an 87-dimensional pass embedding.
In \Cref{fig:model_architecture}(b), the ranker model fuses the 270-dimensional circuit feature with the 87-dimensional pass embedding and projects them into a 256-dimensional circuit representation. Each candidate pass sequence is also encoded into a 256-dimensional sequence representation by a GRU encoder. A cross-attention layer then models the interaction between the circuit representation and each sequence representation, followed by mean pooling and four FC layers to produce a ranking score for each candidate sequence.

In \Cref{fig:model_architecture}(c), we use FAISS for efficient inner product similarity search~\cite{douze2024faisslibrary} over 8,111 circuits based on pass embeddings, retrieving top-$k$ most similar circuits ($k=10$), and use their corresponding optimized pass sequences as candidates. In addition, for each circuit, we precompute the optimization effects of its retrieved top-$k$ candidate pass sequences by compiling them individually and recording each metric, which further accelerates the training of the ranker model.

For training objectives, the pruning model is trained with MSE loss between the predicted pass embeddings and the ground-truth pass embeddings. For the ranker model, we formulate candidate selection as a $k$-way classification problem. Given the predicted ranking scores of the $k$ candidate pass sequences, we use the candidate with the highest ground-truth optimization score as the target label and compute the cross-entropy loss over the predicted scores. During training, we split the 8,111 circuits into 95\% for training and 5\% for validation for both models. The pruning model uses a learning rate of $2\times10^{-3}$ and weight decay of $1\times10^{-2}$, while the ranker model uses a learning rate of $1\times10^{-4}$ and weight decay of $1\times10^{-3}$. Both models use a maximum input length of 2,048 tokens, a batch size of 128, and 200 training epochs.

\subsection{Online Stage}\label{sec:imp_same_online}
We first extract circuit features from the input circuit and use the pruning model to retrieve the top-$k$ candidate sequences ($k=10$) from the optimization dataset, which are then scored by the ranker model.
Based on the ranking results, we perform lightweight BO tuning based on the ranking: pass sequence candidates are evaluated in descending order, and the first valid sequence, i.e., the first sequence that is successfully compiled and passes the hardware validity checks described in \Cref{sec:imp_dataset}, is selected as the initial seed. If all fail, we fall back to the Qiskit $O_3$ sequence. From this seed, we generate 10 initial samples via sequence perturbations (insertion, deletion, replacement), followed by 20 BO iterations with early stopping at 5. The resulting optimized pass sequence is the final output.

%% file: ch5-evaluation.tex
\section{Evaluation}\label{sec:evaluation}

We investigate the following research questions (RQs):

\begin{itemize}[leftmargin=*]
    \item \textbf{RQ1:} Can \system achieve better optimization performance compared to existing baselines? (\Cref{sec:rq1})

    \item \textbf{RQ2:} Does \system maintain consistent performance across unseen quantum algorithm families and backends with different qubit capacities? (\Cref{sec:rq2})
    
    \item \textbf{RQ3:} Does each component of \system contribute effectively to the overall performance? (\Cref{sec:rq3})
    
    \item \textbf{RQ4:} How sensitive is \system to key hyperparameters and dataset size? (\Cref{sec:rq4})

   \item \textbf{RQ5:} Can \system be adapted to different quantum compilers and tuning objectives? (\Cref{sec:rq5})
\end{itemize}

\input{Table/benchmarks}

\input{Table/baseline_settings}

\begin{figure*}[t]
\includegraphics[width=\linewidth]{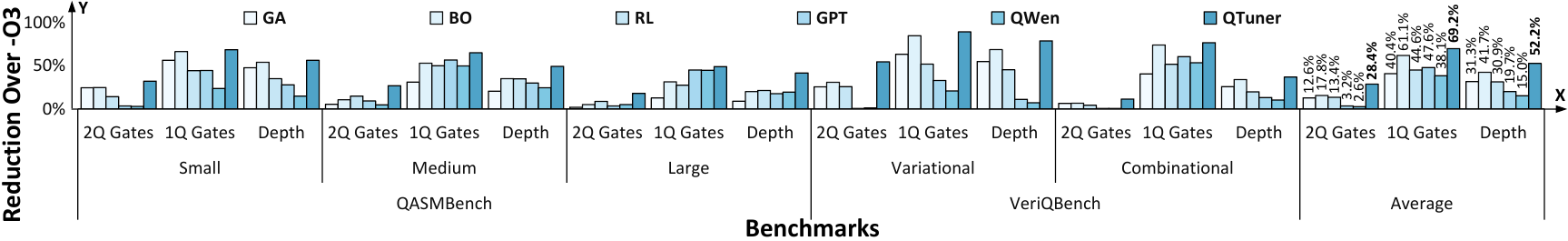}
\caption{Comparison of tuning effectiveness between \system and baselines in Qiskit V2.2.3.}
\label{fig:same_main}
\vspace{-.ex}
\end{figure*}

\begin{figure*}[t]
\includegraphics[width=\linewidth]{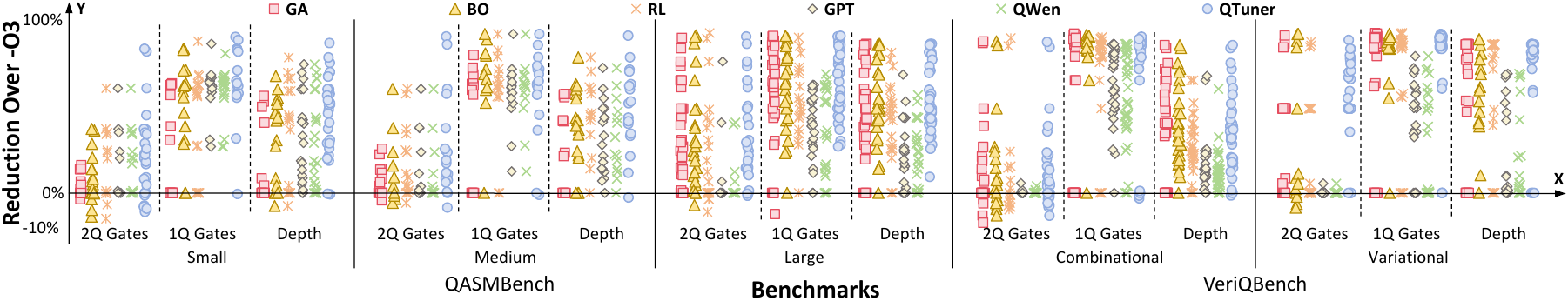}
\caption{Per-case optimization distribution of \system and baselines. A slight x-axis \textbf{\emph{jitter}} improves visibility of overlapping points. Many RL, GPT, and QWen cases concentrate at 0\% reduction because of failed compilations, as described in \Cref{sec:eva_settings}.}
\label{fig:distribution}
\vspace{-2.ex}
\end{figure*}

\subsection{Evaluation Settings}\label{sec:eva_settings}

\noindent
\textbf{Benchmarks}.
We select two open-source benchmarks, QASMBench~\cite{QASMBench} and VeriQBench~\cite{chen2022veriqbenchbenchmarkmultipletypes} (\Cref{table:benchmark_info}), both based on OpenQASM and are widely used in prior studies~\cite{ISCA_qasmbench1,asplos_qasmbench3,PLDI_qasmbench4}. From QASMBench, we select all circuits with no more than 127 qubits, covering representative quantum algorithms. From VeriQBench, we include variational and combinational circuits under the same 127-qubit limit, complementing QASMBench with variational and random circuits.

We convert all OpenQASM 3.0 circuits to 2.0 using the tool described in \Cref{sec:imp_dataset}. To avoid data leakage, we ensure that no circuit file in the selected benchmarks has exactly the same OpenQASM code as any circuit file in our optimization dataset. In total, we use 180 circuits for evaluation.

\noindent
\textbf{Evaluation Metrics}.
Following prior works~\cite{quantum_tuning_cibda,mills2026reinforcementlearningadaptivecomposition}, we compare the numbers of 2Q gates, 1Q gates, and circuit depth against those produced by the Qiskit $O_3$ baseline, and report their reductions relative to the baseline. We impose a 300-second compilation timeout for each evaluated pass sequence. Sequences that fail to compile or exceed this time limit are assigned 0\% reduction across all three evaluation metrics.

\noindent
\textbf{Baselines}.
We consider several representative baselines based on existing traditional and quantum compiler tuning works, including heuristic-based methods such as BO~\cite{boca,cobyan,PDCAT} and GA~\cite{gene1,gene2}, an RL approach based on prior quantum compiler tuning work ~\cite{mqtpredictor,DAC23,mills2026reinforcementlearningadaptivecomposition,quantum_tuning_cibda}, and LLM-based approaches inspired by works of using LLMs for traditional compiler tuning~\cite{pan2026eccoevidencedrivencausalreasoning,qiu2026passbypassoptimizationintentdrivenir}. Detailed configurations are provided in \Cref{table:baseline_setting}, with all random seeds set to 42.

We evaluate all search-based baselines under practical online tuning budgets. In particular,
the BO procedure used for optimization-dataset construction (\Cref{sec:imp_dataset}) allows up to 500 iterations and is intended for offline data generation only; applying the same budget at the evaluation time would incur substantial tuning time overhead.

For BO, the offline dataset-construction runs use early stopping after 50 consecutive non-improving iterations, and we observe that optimization typically saturates well before exhausting the 50-iteration budget. We therefore use a practical budget of 50 iterations for the online BO baseline to balance optimization effectiveness and tuning cost.

For RL, we follow prior PPO-based designs~\cite{mqtpredictor,DAC23}, while strengthening the baseline by expanding its training set from approximately 500 circuits to 4,000 circuits and increasing the RL training budget from 100K to 2M steps. We formulate pass tuning as a Markov decision process. The state consists of static circuit features, normalized circuit metrics, the current step index, and selected-pass counts. The action space contains 29 optimization passes plus \textsc{STOP}, and the terminal reward is the weighted reduction over $O_3$ defined in \Cref{eq:weighted_reduction}, together with step and failure penalties.

For GA, since no directly transferable configuration has been established for quantum compiler optimization pass tuning, we use the same conventional configuration for all circuits, with balanced crossover and mutation rates and moderate selection pressure. The detailed hyperparameters of the GA baseline is reported in \Cref{table:baseline_setting}.

For the LLM baseline, we construct CoT+RAG prompts to guide LLMs in generating optimized pass sequences, inspired by prior work on classical compiler tuning~\cite{qiu2026passbypassoptimizationintentdrivenir}. Specifically, the CoT process instructs the LLM to first analyze the input circuit, features, pass documentation and tuning objective (\Cref{eq:weighted_reduction}), then learn optimization patterns from the RAG-retrieved examples, and finally generate an optimized pass sequence. The prompt includes the input OpenQASM 2.0 circuit, four types of static circuit features, Qiskit optimization pass documentation, and the top-5 retrieved examples. For RAG, we retrieve the top-5 similar circuits from the optimization dataset using inner-product similarity over mean-pooled static circuit features, and use their optimized pass sequences as examples.

\noindent
\textbf{LLMs}.
We select GPT-OSS-120B~\cite{openai2025gptoss120bgptoss20bmodel} as a general-purpose LLM, and QWen3-Coder-480B~\cite{yang2025qwen3technicalreport} as a code-oriented LLM as baselines. To ensure consistency across LLMs, we limit all prompts to 128K tokens, which is the maximum length for GPT-OSS-120B. Since the evaluation prompts require full OpenQASM 2.0 circuit input, we preserve the prompt template and other inputs, and truncate only the OpenQASM 2.0 content when the total length exceeds the 128K limit.

\noindent
\textbf{Evaluation Platforms}.
Our experiments are conducted on a server equipped with a 64-core Intel Xeon Gold CPU and one NVIDIA H100 GPU (80GB memory). For LLMs, we perform inference using Eigen AI~\cite{Web:eigen_ai}. 
For both LLMs, we set the temperature to 0.0, the seed to 42, and the maximum output length to 1,024. Setting the temperature to 0.0 and the seed to 42 helps improve the reproducibility of our experiments.

\input{Table/same_main_time}

\subsection{Main Tuning Results (RQ.1)}\label{sec:rq1}
As shown in \Cref{fig:same_main}, \system achieves higher reduction percentages over the $O_3$ baseline across all three metrics compared to other baselines. Specifically, compared with the strongest baseline (BO), \system achieves absolute improvements of 10.6, 8.1, and 10.5 percentage points in 2Q-gate, 1Q-gate, and circuit depth reduction, respectively. \system also consistently surpasses the other four baselines.

In \Cref{fig:distribution}, we further report the distribution of the per-case optimization effectiveness relative to Qiskit $O_3$. Compared with the GA, BO, and RL baselines, \system has fewer cases below $O_3$ and shows smaller degradation when underperforming. It also produces more cases that outperform $O_3$, with larger metric reductions than all five baselines, indicating that \system more effectively identifies beneficial pass sequences. For RL, GPT, and QWen, many cases concentrate at 0\% because their generated pass sequences often \emph{fail to compile}, whose reductions are set to 0\% as described in \Cref{sec:eva_settings}.

For time efficiency, we report online inference and tuning time only. 
As shown in \Cref{table:main_time}, \system requires an average of 116.8 seconds, reducing the time cost by 73.59\% compared to BO. Although \system is slower than RL (18.3 seconds on average) and LLM-based methods (less than 2 seconds), it achieves significantly better reductions across all three metrics.

Further analysis shows that 11.67\% of the 180 cases (21 cases) trigger the fallback mechanism because all candidate pass sequences produced by the two-stage selection models are invalid. This indicates that the two-stage selection models can identify valid pass sequences for most cases, while the fallback mechanism remains necessary for robustness.

\begin{figure}[t]
\includegraphics[width=\linewidth]{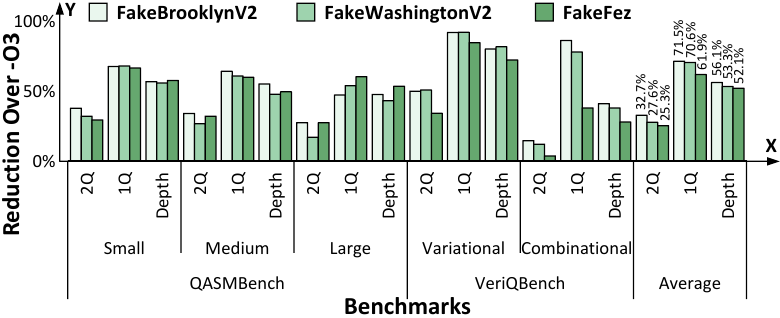}
\caption{Scalability analysis across different quantum backends.}
\label{fig:same_scalability}
\vspace{-1.ex}
\end{figure}

\subsection{Scalability Analysis of \system (RQ.2)}\label{sec:rq2}

\input{Table/sacla_diff_algo}

In this section, we evaluate \system's generalizability and scalability along two dimensions: unseen quantum algorithm families and new backend devices.

\textbf{Generalization to unseen algorithm families.}
Although we ensure that no circuit in the training set is textually identical to any circuit in the evaluation benchmarks, as detailed in \Cref{sec:eva_settings}, quantum programs typically belong to a limited set of representative algorithm families. Therefore, circuits from the same family may still appear in both the training and evaluation sets, but with different number of qubits.

To further examine cross-family generalization, we match circuits by their algorithm-family names and identify 60 benchmark cases whose families are completely absent from the training set. All 60 cases are from QASMBench. As shown in \Cref{tab:family_disjoint}, \system consistently outperforms Qiskit $O_3$ across small, medium, and large circuits. On average, \system reduces the numbers of 2Q gates and 1Q gates, as well as circuit depth, by 23.2\%, 60.7\%, and 48.7\%, respectively. These results show that \system remains effective on quantum algorithm families disjoint from the training set.

\textbf{Scalability to new backend devices.}
As introduced in \Cref{sec:imp_dataset}, the optimization dataset used for pass retrieval in \system is constructed on the 127-qubit FakeWashingtonV2 backend. To evaluate whether the learned optimization knowledge transfers across devices, we further evaluate \system on two unseen backends with different scales: FakeBrooklynV2 (65 qubits) and FakeFez (156 qubits). For a consistent comparison across backends, we use the benchmark circuits in \Cref{table:benchmark_info} containing no more than 65 qubits, such that all selected circuits can be executed on both devices.

As shown in \Cref{fig:same_scalability}, \system achieves better reductions on the FakeBrooklynV2 backend than on FakeWashingtonV2 across all three metrics. On the FakeFez backend, the reduction effectiveness decreases, but the gaps remain small, with only a 2.37\% gap on 2Q gates, the most weighted metric. This suggests that \system remains effective across different backend sizes, although its effectiveness slightly decreases on larger backends. A possible reason is that larger backends introduce a larger layout and routing search space, making pass sequences learned from one backend slightly less transferable.

Nevertheless, fake backends with more than 127 qubits are limited in practice. Among all 67 IBM fake backends~\cite{Web:fakebackends}, only six have 156 qubits and one has 133 qubits, while the rest 60 have no more than 127 qubits. Therefore, the small performance gaps across backend sizes indicate that \system has good scalability for most available fake backends.

\input{Table/ablation_input_feature}

\begin{figure}[t]
\includegraphics[width=\linewidth]{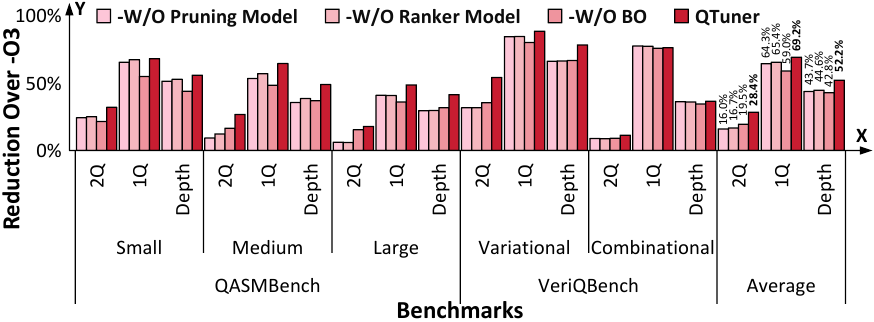}
\caption{Ablation study  of the \system workflow.}
\label{fig:same_ablation}
\vspace{-2.ex}
\end{figure}

\subsection{Ablation Study (RQ.3)}\label{sec:rq3}
The ablation study aims to validate the design of \system from two aspects: (1) input feature selection, (2) the workflow including the two models and lightweight BO refinement.

\subsubsection{Input Feature}
As shown in \Cref{table:ablation_input_feature}, we perform feature ablation by removing each of the four circuit feature types for the pruning model, and circuit features and pass embeddings separately for the ranker model. Results on the validation set show that removing any feature degrades performance, indicating that all features contribute positively. Therefore, all features are retained in the final design.

\subsubsection{Overall Workflow}
We conduct ablation studies by separately removing each component of \system: the pruning model, the ranker model, and lightweight BO. To keep the pipeline functional after removing each component, we replace the pruning model with circuit-feature-similarity-based retrieval, order candidates by retrieval similarity when removing the ranker, and directly output the first valid ranked sequence when removing lightweight BO. In all settings, $O_3$ is used as the fallback if all candidates fail. As shown in \Cref{fig:same_ablation}, removing any component degrades performance, with 2Q gate reduction dropping by 8.9\% -- 12.4\% in terms of absolute value, which demonstrates the necessity of all components in \system.


\begin{figure*}[t]
\centering
\includegraphics[width=\linewidth]{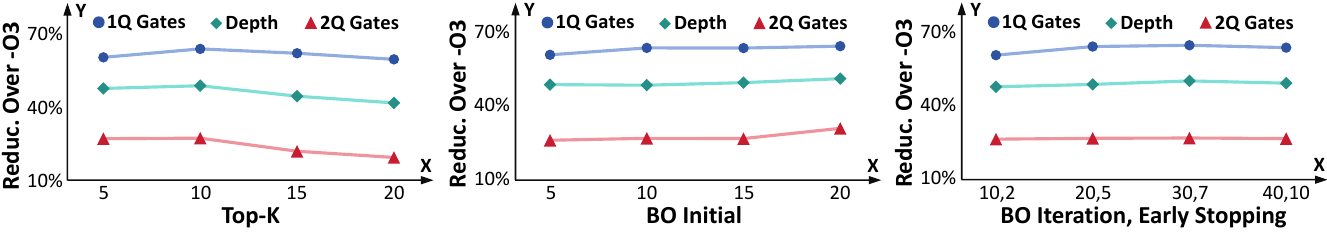}
\caption{Hyperparameter analysis  of \system.}
\label{fig:same_sensitive}
\vspace{-2.ex}
\end{figure*}

\begin{figure}[t]
\includegraphics[width=\linewidth]{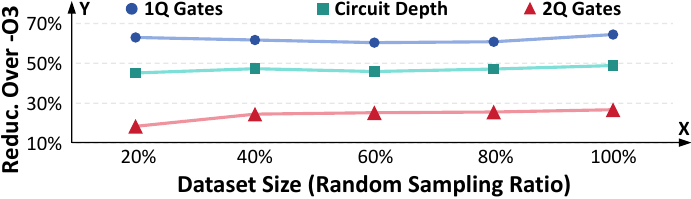}
\caption{Dataset size sampling analysis  of \system.}
\label{fig:dataset_size}
\vspace{-2.ex}
\end{figure}

\subsection{Sensitivity Analysis (RQ.4)}\label{sec:rq4}
We conduct all sensitivity analyses in the following subsections using all 20 cases from \texttt{QASMBench/Medium}.

\subsubsection{Hyperparameter}
The sensitivity evaluation results of the three key hyperparameters are shown in \Cref{fig:same_sensitive}.

For \textbf{Top-$k$}, the default setting value is 10. Reducing $k$ to 5 achieves comparable performance on 2Q gates and depth, with only a slight decrease on 1Q gates. Increasing $k$ to 15 or 20 leads to minor degradation, because \system uses the first valid ranked sequence as the initial seed for lightweight BO and falls back to the default $O_3$ sequence if all candidates fail (\Cref{sec:imp_same_online}). A larger $k$ may introduce lower-quality candidates, increasing the risk of suboptimal BO initialization. Overall, the small performance variation indicates that \system is not highly sensitive to $k$.

For \textbf{BO initial samples}, increasing the number of initial samples slightly improves the reduction of all metrics due to a larger search space, but also increases time cost. We therefore choose 10 as a balance between efficiency and performance.

For \textbf{BO iterations}, we vary the iteration count in steps of 10 and adjust early stopping accordingly. Reduction improves from (iteration=10, early stopping=2) to (20, 5), but shows no significant gain beyond (30, 7) or (40, 10). This indicates that \system already provides strong initialization, allowing BO to converge within a small number of iterations. Thus, we select 20 iterations with early stopping at 5 as a good trade-off.

\subsubsection{Dataset Size}\label{sec:dataset_size}

We further evaluate the sensitivity of \system to the optimization dataset size by randomly sampling 20\%, 40\%, 60\%, 80\%, and 100\% of the dataset. As shown in \Cref{fig:dataset_size}, the 2Q gate reduction improves noticeably from 20\% to 40\%, while the gains become marginal beyond 40\%. This indicates that \system becomes stable once the dataset reaches 40\%, as this subset already covers representative optimized pass sequences for diverse quantum circuits.


\begin{figure}[t]
\includegraphics[width=\linewidth]{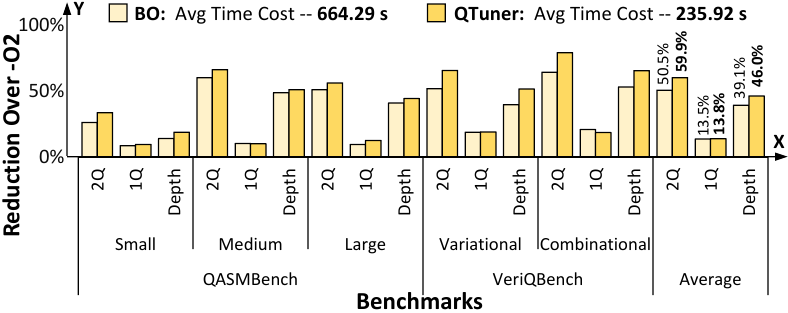}
\caption{Comparison between \system and BO in PyTKET V2.13.0 over its highest optimization level $O_2$.}
\label{fig:pytket}
\vspace{-2.ex}
\end{figure}

\subsection{Adaptability of \system (RQ.5)}\label{sec:rq5}

\subsubsection{Adaptation to a Different Quantum Compiler}\label{sec:pytket}

We conduct an adaptability experiment on the PyTKET compiler V2.13.0. We include all 22 PyTKET optimization passes as the tuning search space and use the highest optimization level $O_2$ of PyTKET as the reduction baseline. We still use IBM FakeWashingtonV2 as the target backend device for both dataset  construction and evaluation, accessed through the \texttt{IBMQBackend} interface in \texttt{pytket-qiskit} v0.77.0.

We continue to use QCircuitBench to construct the PyTKET-specific optimization dataset, and apply the long-term BO optimization with the same BO settings as in \Cref{sec:imp_dataset}. We then profile PyTKET-specific pass embeddings, and train the pruning and ranker models from scratch following the same workflow as in the Qiskit compiler.

For evaluation, we compare \system with BO, the strongest baseline in \Cref{sec:rq1}, using the same settings as in \Cref{table:baseline_setting}. In \Cref{fig:pytket}, \system achieves better reduction effectiveness, achieving an average improvement of 9.4 percentage points in 2Q-gate reduction over BO while reducing tuning time by 64.49\%. Notably, PyTKET shows larger 2Q gate reductions but smaller 1Q gate reductions than Qiskit. This may be because PyTKET's default $O_2$ optimization already performs strong 1Q gate simplification and optimization, whereas 2Q gates still benefit from pass tuning.

\begin{figure}[t]
\includegraphics[width=\linewidth]{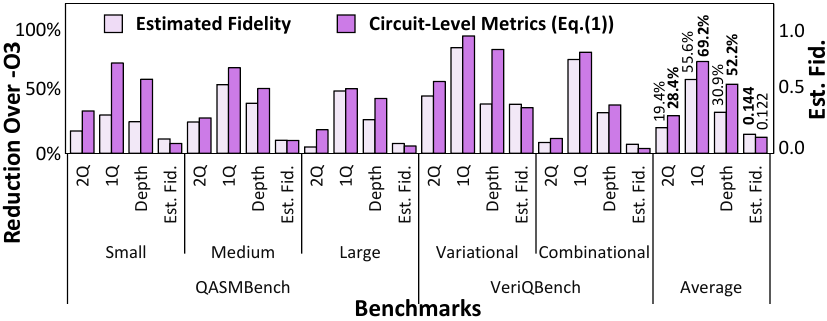}
\caption{Comparison of estimated fidelity (EF) and circuit-level metrics (\Cref{eq:weighted_reduction}) as tuning objectives.}
\label{fig:efp}
\vspace{-2.ex}
\end{figure}

\subsubsection{Adaptation to a Different Tuning Objective}\label{sec:efp}

As discussed in \Cref{sec:tuning_obj}, our main experiments use circuit-level metrics as the tuning objective. Here, we further adapt \system to an estimated-fidelity-oriented objective. Following prior work~\cite{mqtpredictor,tuniq}, estimated fidelity (EF) is computed as the product of the expected fidelities of all gates on the target backend. Specifically, we still use the FakeWashingtonV2 backend and reconstruct the optimization dataset using EF as the objective, while keeping the BO settings consistent with \Cref{sec:imp_dataset}. During pass profiling, we represent the pass-induced relative change as $\Delta_{\mathrm{EF}}$, retrain two models, and use EF as the objective for lightweight BO refinement.

In \Cref{fig:efp}, we compare the two objectives using circuit-level reductions and average EF. We report EF as an absolute value because complex circuits often have very low EF (e.g., 0.001), where small absolute gains may lead to overly large relative improvements. The results show that the circuit-level objective achieves larger reductions in gate counts and depth, while the EF objective achieves higher average EF by explicitly targeting gates with higher backend-specific error rates.

Meanwhile, the circuit-level objective achieves an average EF only 0.022 lower in absolute value, indicating that reducing overall 1Q and 2Q gate counts can also effectively improve EF. These results show that \system can be adapted to different optimization objectives by rerunning the overall workflow with the corresponding optimization objective.

\subsection{Limitations}

One limitation of \system is that its offline stage must be reconstructed when adapting to a new compiler or tuning objective. However, this process is performed only once offline, and the resulting dataset and models can be reused for future circuits. Compared with per-circuit exhaustive tuning, \system still substantially reduces online tuning cost by retrieving and refining promising pass sequences.

%% file: Table/benchmarks.tex
\begin{table}[t]
\caption{Selected OpenQASM benchmarks in evaluation.}
\begin{adjustbox}{max width=\linewidth}
\setlength{\tabcolsep}{0.7ex}
\setstretch{1.}
\label{table:benchmark_info}
\begin{tabular}{ccccl}
\hline
\textbf{Benchmark}                   & \textbf{Size}          & \textbf{\#Cases} & \textbf{Qubits}    & \multicolumn{1}{c}{\textbf{Description}}                                                                                \\ \hline
\multirow{3}{*}{QASMBench}  & Small         & 38   & 3 -- 10   & \multirow{3}{*}{\begin{tabular}[c]{@{}l@{}}Quantum algorithms\end{tabular}}                                            \\ \cline{2-4}
                            & Medium        & 20   & 11 -- 27  &                                                                                                                \\ \cline{2-4}
                            & Large         & 33   & 28 -- 118 &                                                                                                                \\ \hline
\multirow{2}{*}{VeriQBench} & Variational   & 47   & 4 -- 121  & Variational circuits \\ \cline{2-5} 
                            & Combinational & 42   & 1 -- 100  & Random circuits                                                                   \\ \hline
\multicolumn{2}{c}{Total}                   & \multicolumn{3}{c}{180}                                                                                                           \\ \hline
\end{tabular}
\end{adjustbox}
\vspace{-2.ex}
\end{table}

%% file: Table/baseline_settings.tex
\begin{table*}[t]
\caption{Hyperparameter settings for different baselines in the evaluation.}
\begin{adjustbox}{max width=\linewidth}
\setlength{\tabcolsep}{.3ex}
\setstretch{1.}
\label{table:baseline_setting}
\begin{tabular}{lll}
\hline
\textbf{Methods}                & \textbf{Abbr.} & \textbf{Settings} \\ \hline
\begin{tabular}[c]{@{}l@{}}Bayesian \\Optimization\end{tabular}  & BO           & \begin{tabular}[c]{@{}l@{}}Initial samples: 10, iterations: 50, early stopping: 20, other settings are detailed in \Cref{sec:imp_dataset}. \end{tabular}  \\ \hline
\begin{tabular}[c]{@{}l@{}}Genetic \\Algorithm\end{tabular}      & GA           & \begin{tabular}[c]{@{}l@{}}Initial samples: 20, iterations: 50, population size: 8,  generation size: 12, elite size: 4,   crossover rate: 0.5, \\ mutation rate: 0.5, keep\_improved: 5, tournament size:3\end{tabular} \\ \hline
\begin{tabular}[c]{@{}l@{}}Reinforcement\\ Learning\end{tabular} & RL           & \begin{tabular}[c]{@{}l@{}}Train a PPO agent on 4,000  random samples  from  the \\ QCircuitBench with  2,000,000 steps,  batch   size 2048, learning rate $1\mathrm{e}{-4}$,  clip\_range 0.2, step penalty 0.015\end{tabular} \\ \hline
LLMs          & \begin{tabular}[c]{@{}l@{}} GPT  \\QWen  \end{tabular}       & {\begin{tabular}[c]{@{}l@{}}CoT+RAG Prompts; RAG: retrieving top-5 optimized pass sequences;  CoT: guides the LLM to analyze  the circuit, \\ features, pass documentation, tuning objectives, and examples  before generating a pass sequence \end{tabular}} \\ \hline
\multicolumn{2}{l}{\system}                                 & {\begin{tabular}[c]{@{}l@{}} Top-K sample: 10, BO Initial samples: 10, iteration: 20, early stopping: 5 \end{tabular}} \\ \hline
\end{tabular}
\end{adjustbox}
\vspace{-1.ex}
\end{table*}

%% file: Table/same_main_time.tex
\begin{table}[t]
\caption{Online tuning time comparison on Qiskit.}
\begin{adjustbox}{max width=\linewidth}
\setlength{\tabcolsep}{.5ex}
\setstretch{1.}
\label{table:main_time}
\begin{tabular}{cc|ccccc|c}
\hline
\multicolumn{2}{c|}{\textbf{Benchmark}}                                                                                                       & \textbf{GA}     & \textbf{BO}    & \textbf{RL}   & \textbf{GPT} & \textbf{QWen}\ & \textbf{\system} \\ \hline
\multirow{3}{*}{\begin{tabular}[c]{@{}c@{}}QASM-\\ Bench\end{tabular}}  & Small                                                     & 598.5 s                 & 196.7 s                 & 18.6 s                  & 1.1 s                                                                         & 1.4 s                                                                         & 27.9 s                         \\ \cline{2-8} 
                                                                        & Medium                                                    & 1,652.6 s                & 292.2 s                 & 8.2 s                   & 1.1 s                                                                         & 1.3 s                                                                          & 95.5 s                          \\ \cline{2-8} 
                                                                        & Large                                                     & 1,837.1 s                 & 808.3 s                  & 54.5 s                   & 1.2 s                                                                         & 1.0 s                                                                         & 331.7 s                         \\ \hline
\multirow{2}{*}{\begin{tabular}[c]{@{}c@{}}VeriQ-\\ Bench\end{tabular}} & Variational   & 1,109.7 s                & 379.2 s                  & 3.4 s                    & 1.1 s                                                                        & 1.1 s                                                                          & 53.4 s                          \\ \cline{2-8} 
                                                                        & Combinational & 1,665.6 s                & 535.3 s                 & 6.9 s                   & 1.2 s                                                                         & 1.1 s                                                                         & 75.7 s                         \\ \hline
\multicolumn{2}{c|}{Average}                                                                                                         & 1,372.5 s                & 442.3 s                  & 18.3 s                   & 1.1 s                                                                        & 1.2 s                                                                         & 116.8 s                        \\ \hline
\end{tabular}
\end{adjustbox}
\vspace{-2.ex}
\end{table}

%% file: Table/sacla_diff_algo.tex
\begin{table}[t]
\centering
\caption{\system's reduction over Qiskit $O_3$ on algorithm-family-disjoint cases from QASMBench.}
\label{tab:family_disjoint}
\begin{adjustbox}{max width=\linewidth}
\begin{tabular}{lrrrr}
\toprule
\textbf{QASMBench} & \textbf{\#Cases} & \textbf{2Q Gates} & \textbf{1Q Gates} & \textbf{Depth} \\
\midrule
Small  & 28 & 32.5\% & 68.9\% & 56.8\% \\
Medium & 13 & 23.8\% & 59.6\% & 45.4\% \\
Large  & 19 &  9.3\% & 49.3\% & 39.1\% \\ \hline
Average & 60 &  23.2\% & 60.7\% & 48.7\% \\
\bottomrule
\end{tabular}
\end{adjustbox}
\vspace{-2.ex}
\end{table}

%% file: Table/ablation_input_feature.tex
\begin{table}[t]
\caption{Input-feature ablation study measured by validation loss during training. \emph{Lower is better.}}
\begin{adjustbox}{max width=\linewidth}
\setlength{\tabcolsep}{.8ex}
\setstretch{1.}
\label{table:ablation_input_feature}
\begin{tabular}{rc|rc}
\hline
\multicolumn{1}{c}{\textbf{Model}}         & \textbf{MSE} & \multicolumn{1}{c}{\textbf{Model}}                                                      & \textbf{Cross-Entropy} \\ \hline
\multicolumn{1}{l}{Pruning Model} & \textbf{7.5x$10^{-3}$}                                                     & \multicolumn{1}{l}{Ranker Model}                                               & \textbf{1.5606}                                                                          \\ \hline
-W/O DAG Node                     & 11.5x$10^{-3}$                                                      & \multirow{2}{*}{\begin{tabular}[c]{@{}r@{}}-W/O Static\\ Feature\end{tabular}} & \multirow{2}{*}{1.7881}                                                         \\ \cline{1-2}
-W/O Structure                    & 8.0x$10^{-3}$                                                       &                                                                                &                                                                                  \\ \hline
-W/O Scale                        & 7.7x$10^{-3}$                                                      & \multirow{2}{*}{\begin{tabular}[c]{@{}r@{}}-W/O Pass\\ Embedding\end{tabular}} & \multirow{2}{*}{1.9414}                                                         \\ \cline{1-2}
-W/O Interaction Graph            & 8.5x$10^{-3}$                                                      &                                                                                &                                                                                  \\ \hline
\end{tabular}
\end{adjustbox}
\vspace{-1.ex}
\end{table}

%% file: ch6-related_work.tex
\section{Related Work}

\noindent
\textbf{Compiler Auto-tuning.}
Traditional compiler auto-tuning has explored heuristic search, including BO and GA~\cite{boca,cobyan,PDCAT,gene1,gene2}, RL-based pass selection~\cite{rl1,rl3,rl4,oopsla_phase,autophase}, similarity-based retrieval~\cite{ICS_Similarity,SC_similarity}, and LLM-guided optimization~\cite{pan2026eccoevidencedrivencausalreasoning,qiu2026passbypassoptimizationintentdrivenir,meta_llm_compiler}. Quantum compiler tuning also adopts heuristic and RL-based methods~\cite{mqtpredictor,tuniq,DAC23,quantum_tuning_cibda,mills2026reinforcementlearningadaptivecomposition,dangwal2025cliffordassistedoptimalpass}, but existing studies tune only limited optimization pass subsets. \system instead tunes the full Qiskit and PyTKET optimization-pass space with a similarity-based design.

\noindent
\textbf{Quantum Software Engineering}.
Recent advances in quantum computing have led to increasing interest in quantum software engineering~\cite{qse_roadmap}, encompassing areas such as quantum compiler tuning~\cite{DAC23,mills2026reinforcementlearningadaptivecomposition,dangwal2025cliffordassistedoptimalpass,quantum_tuning_cibda}, execution performance and time prediction~\cite{tosem_circuits,quest}, testing~\cite{qse_test1,qse_test2,qse_test3,qse_test4}, program repair~\cite{qse_repair}, and quantum code generation~\cite{qse_codegen2,ASE_Quanbench}.

%% file: ch7-conclusion.tex
\section{Threats to Validity}

\textbf{Internal Validity.}
A primary threat to internal validity lies in baseline construction. Existing quantum compiler tuning studies do not target the full optimization pass space considered in this work, and some key implementation hyperparameters are not available. Therefore, we re-implement the baselines based on their core designs, which may introduce deviations from the original methods. To mitigate this threat, we preserve their main optimization strategies and adopt commonly used hyperparameter settings when details are missing. Another threat lies in the LLM variance, which is reduced by setting temperature to 0.0 and seed to 42. Finally, our weighted objective may favor 2Q gates optimization, but aligns with prior work emphasizing its importance~\cite{mills2026reinforcementlearningadaptivecomposition,guoq}.

\textbf{External Validity.}
Our evaluation focuses on Qiskit, PyTKET, and IBM fake backends with heavy-hex architectures. Therefore, the results may not fully generalize to other SDKs and hardware topologies. Since our evaluation metrics focus on circuit structural properties, the results are expected to remain comparable as long as the simulator and real quantum machine share the same gate set and coupling topology.

\section{Conclusion}
In this paper, we present \system, a feature-guided AI-driven framework for tuning quantum compiler optimization passes over the full pass space. \system constructs a large-scale optimization dataset, represents circuits with static features and optimization-aware pass embeddings, and uses two models to retrieve and rank promising pass sequences, followed by lightweight BO refinement. Experiments on Qiskit and PyTKET show that \system improves circuit optimization effectiveness while reducing online pass tuning cost.


\section*{Acknowledgment}
We thank all reviewers for their constructive comments.

%% file: bibfile.bib
@String{Computing = "Computing" }

@String{Computer = "{IEEE} Computer" }

@inproceedings{meta_llm_compiler,
author = {Cummins, Chris and Seeker, Volker and Grubisic, Dejan and Roziere, Baptiste and Gehring, Jonas and Synnaeve, Gabriel and Leather, Hugh},
title = {{LLM Compiler: Foundation Language Models for Compiler Optimization}},
year = {2025},
isbn = {9798400714078},
publisher = {Association for Computing Machinery},
address = {New York, NY, USA},
url = {https://doi.org/10.1145/3708493.3712691},
doi = {10.1145/3708493.3712691},
booktitle = {Proceedings of the 34th ACM SIGPLAN International Conference on Compiler Construction},
pages = {141–153},
numpages = {13},
location = {Las Vegas, NV, USA},
series = {CC '25}
}

@article{EditDis,
author = {Wagner, Robert A. and Fischer, Michael J.},
title = {The String-to-String Correction Problem},
year = {1974},
issue_date = {Jan. 1974},
publisher = {Association for Computing Machinery},
address = {New York, NY, USA},
volume = {21},
number = {1},
issn = {0004-5411},
url = {https://doi.org/10.1145/321796.321811},
doi = {10.1145/321796.321811},
journal = {J. ACM},
month = jan,
pages = {168–173},
numpages = {6}
}

@inproceedings{GAForReduceCodeSize,
author = {Cooper, Keith D. and Schielke, Philip J. and Subramanian, Devika},
title = {Optimizing for Reduced Code Space Using Genetic Algorithms},
year = {1999},
isbn = {1581131364},
publisher = {Association for Computing Machinery},
address = {New York, NY, USA},
url = {https://doi.org/10.1145/314403.314414},
doi = {10.1145/314403.314414},
booktitle = {Proceedings of the ACM SIGPLAN 1999 Workshop on Languages, Compilers, and Tools for Embedded Systems},
pages = {1–9},
numpages = {9},
location = {Atlanta, Georgia, USA},
series = {LCTES '99}
}

@inproceedings{boca,
author = {Chen, Junjie and Xu, Ningxin and Chen, Peiqi and Zhang, Hongyu},
title = {{Efficient Compiler Autotuning via Bayesian Optimization}},
year = {2021},
isbn = {9781450390859},
publisher = {IEEE Press},
url = {https://doi.org/10.1109/ICSE43902.2021.00110},
doi = {10.1109/ICSE43902.2021.00110},
booktitle = {Proceedings of the 43rd International Conference on Software Engineering},
pages = {1198–1209},
numpages = {12},
location = {Madrid, Spain},
series = {ICSE '21}
}

@article{PDCAT,
author = {Zhu, Mingxuan and Sun, Zeyu and Hao, Dan},
title = {{PDCAT: Preference-Driven Compiler Auto-tuning}},
year = {2025},
issue_date = {July 2025},
publisher = {Association for Computing Machinery},
address = {New York, NY, USA},
volume = {2},
number = {FSE},
url = {https://doi.org/10.1145/3715756},
doi = {10.1145/3715756},
journal = {Proc. ACM Softw. Eng.},
month = jun,
articleno = {FSE039},
numpages = {21}
}

@article{cobyan,
author = {Zhu, Mingxuan and Hao, Dan and Chen, Junjie},
title = {{Compiler Autotuning through Multiple-phase Learning}},
year = {2024},
issue_date = {May 2024},
publisher = {Association for Computing Machinery},
address = {New York, NY, USA},
volume = {33},
number = {4},
issn = {1049-331X},
url = {https://doi.org/10.1145/3640330},
doi = {10.1145/3640330},
journal = {ACM Trans. Softw. Eng. Methodol.},
month = apr,
articleno = {100},
numpages = {38}
}

@inproceedings{gene1,
author = {Garciarena, Unai and Santana, Roberto},
title = {{Evolutionary Optimization of Compiler Flag Selection by Learning and Exploiting Flags Interactions}},
year = {2016},
isbn = {9781450343237},
publisher = {Association for Computing Machinery},
address = {New York, NY, USA},
url = {https://doi.org/10.1145/2908961.2931696},
doi = {10.1145/2908961.2931696},
booktitle = {Proceedings of the 2016 on Genetic and Evolutionary Computation Conference Companion},
pages = {1159–1166},
numpages = {8},
location = {Denver, Colorado, USA},
series = {GECCO '16 Companion}
}

@inproceedings{gene2,
author = {Pan, Haolin and Wei, Yuanyu and Xing, Mingjie and Wu, Yanjun and Zhao, Chen},
title = {{Towards Efficient Compiler Auto-tuning: Leveraging Synergistic Search Spaces}},
year = {2025},
isbn = {9798400712753},
publisher = {Association for Computing Machinery},
address = {New York, NY, USA},
url = {https://doi.org/10.1145/3696443.3708961},
doi = {10.1145/3696443.3708961},
booktitle = {Proceedings of the 23rd ACM/IEEE International Symposium on Code Generation and Optimization},
pages = {614–627},
numpages = {14},
location = {Las Vegas, NV, USA},
series = {CGO '25}
}

@inproceedings{rl1,
author = {Coons, Katherine E. and Robatmili, Behnam and Taylor, Matthew E. and Maher, Bertrand A. and Burger, Doug and McKinley, Kathryn S.},
title = {{Feature selection and policy optimization for distributed instruction placement using reinforcement learning}},
year = {2008},
isbn = {9781605582825},
publisher = {Association for Computing Machinery},
address = {New York, NY, USA},
url = {https://doi.org/10.1145/1454115.1454122},
doi = {10.1145/1454115.1454122},
booktitle = {Proceedings of the 17th International Conference on Parallel Architectures and Compilation Techniques},
pages = {32–42},
numpages = {11},
location = {Toronto, Ontario, Canada},
series = {PACT '08}
}

@inproceedings{rl3,
author = {Cummins, Chris and Wasti, Bram and Guo, Jiadong and Cui, Brandon and Ansel, Jason and Gomez, Sahir and Jain, Somya and Liu, Jia and Teytaud, Olivier and Steiner, Benoit and Tian, Yuandong and Leather, Hugh},
title = {{CompilerGym: robust, performant compiler optimization environments for AI research}},
year = {2022},
isbn = {9781665405843},
publisher = {IEEE Press},
url = {https://doi.org/10.1109/CGO53902.2022.9741258},
doi = {10.1109/CGO53902.2022.9741258},
booktitle = {Proceedings of the 20th IEEE/ACM International Symposium on Code Generation and Optimization},
pages = {92–105},
numpages = {14},
location = {Virtual Event, Republic of Korea},
series = {CGO '22}
}

@inproceedings{rl4,
author = {Park, Sunghyun and Latifi, Salar and Park, Yongjun and Behroozi, Armand and Jeon, Byungsoo and Mahlke, Scott},
title = {{SRTuner: effective compiler optimization customization by exposing synergistic relations}},
year = {2022},
isbn = {9781665405843},
publisher = {IEEE Press},
url = {https://doi.org/10.1109/CGO53902.2022.9741263},
doi = {10.1109/CGO53902.2022.9741263},
booktitle = {Proceedings of the 20th IEEE/ACM International Symposium on Code Generation and Optimization},
pages = {118–130},
numpages = {13},
location = {Virtual Event, Republic of Korea},
series = {CGO '22}
}

@article{pan2026eccoevidencedrivencausalreasoning,
  title   = {{ECCO: Evidence-Driven Causal Reasoning for Compiler Optimization}},
  author  = {Pan, Haolin and Huang, Lianghong and Dong, Jinyuan and Xing, Mingjie and Wu, Yanjun},
  journal = {arXiv preprint arXiv:2602.00087},
  year    = {2026},
  doi     = {10.48550/arXiv.2602.00087},
  url     = {https://arxiv.org/abs/2602.00087}
}

@inproceedings{
yang2025qcircuitbench,
title={{QCircuitBench: A Large-Scale Dataset for Benchmarking Quantum Algorithm Design}},
author={Rui Yang and Ziruo Wang and Yuntian Gu and Yitao Liang and Tongyang Li},
booktitle={The Thirty-ninth Annual Conference on Neural Information Processing Systems Datasets and Benchmarks Track},
year={2025},
url={https://openreview.net/forum?id=NkiLldW2bi}
}

@article{openai2025gptoss120bgptoss20bmodel,
      title={{gpt-oss-120b \& gpt-oss-20b Model Card}}, 
      author={Sandhini Agarwal and Lama Ahmad and Jason Ai and Sam Altman and Andy Applebaum and Edwin Arbus and Rahul K. Arora and Yu Bai and Bowen Baker and Haiming Bao and Boaz Barak and Ally Bennett and Tyler Bertao and Nivedita Brett and Eugene Brevdo and Greg Brockman and Sebastien Bubeck and Che Chang and Kai Chen and Mark Chen and Enoch Cheung and Aidan Clark and Dan Cook and Marat Dukhan and Casey Dvorak and Kevin Fives and Vlad Fomenko and Timur Garipov and Kristian Georgiev and Mia Glaese and Tarun Gogineni and Adam Goucher and Lukas Gross and Katia Gil Guzman and John Hallman and Jackie Hehir and Johannes Heidecke and Alec Helyar and Haitang Hu and Romain Huet and Jacob Huh and Saachi Jain and Zach Johnson and Chris Koch and Irina Kofman and Dominik Kundel and Jason Kwon and Volodymyr Kyrylov and Elaine Ya Le and Guillaume Leclerc and James Park Lennon and Scott Lessans and Mario Lezcano-Casado and Yuanzhi Li and Zhuohan Li and Ji Lin and Jordan Liss and Lily and Liu and Jiancheng Liu and Kevin Lu and Chris Lu and Zoran Martinovic and Lindsay McCallum and Josh McGrath and Scott McKinney and Aidan McLaughlin and Song Mei and Steve Mostovoy and Tong Mu and Gideon Myles and Alexander Neitz and Alex Nichol and Jakub Pachocki and Alex Paino and Dana Palmie and Ashley Pantuliano and Giambattista Parascandolo and Jongsoo Park and Leher Pathak and Carolina Paz and Ludovic Peran and Dmitry Pimenov and Michelle Pokrass and Elizabeth Proehl and Huida Qiu and Gaby Raila and Filippo Raso and Hongyu Ren and Kimmy Richardson and David Robinson and Bob Rotsted and Hadi Salman and Suvansh Sanjeev and Max Schwarzer and D. Sculley and Harshit Sikchi and Kendal Simon and Karan Singhal and Yang Song and Dane Stuckey and Zhiqing Sun and Philippe Tillet and Sam Toizer and Foivos Tsimpourlas and Nikhil Vyas and Eric Wallace and Xin Wang and Miles Wang and Olivia Watkins and Kevin Weil and Amy Wendling and Kevin Whinnery and Cedric Whitney and Hannah Wong and Lin Yang and Yu Yang and Michihiro Yasunaga and Kristen Ying and Wojciech Zaremba and Wenting Zhan and Cyril Zhang and Brian Zhang and Eddie Zhang and Shengjia Zhao},
      journal = {arXiv preprint arXiv:2508.10925},
      year = {2025},
      doi     = {10.48550/arXiv.2508.10925},
      url     = {https://arxiv.org/abs/2508.10925}
}

@article{yang2025qwen3technicalreport,
  title   = {{Qwen3 Technical Report}},
  author  = {Yang, An and Li, Anfeng and Yang, Baosong and Zhang, Beichen and Hui, Binyuan and Zheng, Bo and Yu, Bowen and Gao, Chang and Huang, Chengen and Lv, Chenxu and Zheng, Chujie and Liu, Dayiheng and Zhou, Fan and Huang, Fei and Hu, Feng and Ge, Hao and Wei, Haoran and Lin, Huan and Tang, Jialong and Yang, Jian and Tu, Jianhong and Zhang, Jianwei and Yang, Jianxin and Yang, Jiaxi and Zhou, Jing and Zhou, Jingren and Lin, Junyang and Dang, Kai and Bao, Keqin and Yang, Kexin and Yu, Le and Deng, Lianghao and Li, Mei and Xue, Mingfeng and Li, Mingze and Zhang, Pei and Wang, Peng and Zhu, Qin and Men, Rui and Gao, Ruize and Liu, Shixuan and Luo, Shuang and Li, Tianhao and Tang, Tianyi and Yin, Wenbiao and Ren, Xingzhang and Wang, Xinyu and Zhang, Xinyu and Ren, Xuancheng and Fan, Yang and Su, Yang and Zhang, Yichang and Zhang, Yinger and Wan, Yu and Liu, Yuqiong and Wang, Zekun and Cui, Zeyu and Zhang, Zhenru and Zhou, Zhipeng and Qiu, Zihan},
  journal = {arXiv preprint arXiv:2505.09388},
  year    = {2025},
  doi     = {10.48550/arXiv.2505.09388},
  url     = {https://arxiv.org/abs/2505.09388}
}

@article{dangwal2025cliffordassistedoptimalpass,
  title   = {{Clifford Assisted Optimal Pass Selection for Quantum Transpilation}},
  author  = {Dangwal, Siddharth and Ravi, Gokul Subramanian and Seifert, Lennart Maximilian and Das, Poulami and Sud, James and Chong, Frederic T.},
  journal = {arXiv preprint arXiv:2306.15020},
  year    = {2023},
  doi     = {10.48550/arXiv.2306.15020},
  url     = {https://arxiv.org/abs/2306.15020}
}

@article{mqtpredictor,
author = {Quetschlich, Nils and Burgholzer, Lukas and Wille, Robert},
title = {{MQT Predictor: Automatic Device Selection with Device-Specific Circuit Compilation for Quantum Computing}},
year = {2025},
issue_date = {March 2025},
publisher = {Association for Computing Machinery},
address = {New York, NY, USA},
volume = {6},
number = {1},
url = {https://doi.org/10.1145/3673241},
doi = {10.1145/3673241},
journal = {ACM Transactions on Quantum Computing},
month = jan,
articleno = {10},
numpages = {26}
}

@inproceedings{guoq,
author = {Xu, Amanda and Molavi, Abtin and Tannu, Swamit and Albarghouthi, Aws},
title = {{Optimizing Quantum Circuits, Fast and Slow}},
year = {2025},
isbn = {9798400706981},
publisher = {Association for Computing Machinery},
address = {New York, NY, USA},
url = {https://doi.org/10.1145/3669940.3707240},
doi = {10.1145/3669940.3707240},
booktitle = {Proceedings of the 30th ACM International Conference on Architectural Support for Programming Languages and Operating Systems, Volume 1},
pages = {777–793},
numpages = {17},
location = {Rotterdam, Netherlands},
series = {ASPLOS '25}
}

@INPROCEEDINGS{DAC23,
  author={Quetschlich, Nils and Burgholzer, Lukas and Wille, Robert},
  booktitle={2023 60th ACM/IEEE Design Automation Conference (DAC)}, 
  title={{Compiler Optimization for Quantum Computing Using Reinforcement Learning}}, 
  year={2023},
  volume={},
  number={},
  pages={1-6},
  doi={10.1109/DAC56929.2023.10248002}
}

@article{mills2026reinforcementlearningadaptivecomposition,
  title   = {{Reinforcement Learning for Adaptive Composition of Quantum Circuit Optimisation Passes}},
  author  = {Mills, Daniel and Williams, Ifan and Swain, Jacob and Matos, Gabriel and Rinaldi, Enrico and Koziell-Pipe, Alexander},
  journal = {arXiv preprint arXiv:2601.21629},
  year    = {2026},
  doi     = {10.48550/arXiv.2601.21629},
  url     = {https://arxiv.org/abs/2601.21629}
}

@article{inter_graph,
doi = {10.1088/1367-2630/ad264d},
url = {https://doi.org/10.1088/1367-2630/ad264d},
year = {2024},
month = {feb},
publisher = {IOP Publishing},
volume = {26},
number = {2},
pages = {023033},
author = {Kyprianidis, Antonis and Rasmusson, A J and Richerme, Philip},
title = {Interaction graph engineering in trapped-ion quantum simulators with global drives},
journal = {New Journal of Physics}
}

@INPROCEEDINGS{supermarq,
  author={Tomesh, Teague and Gokhale, Pranav and Omole, Victory and Ravi, Gokul Subramanian and Smith, Kaitlin N. and Viszlai, Joshua and Wu, Xin-Chuan and Hardavellas, Nikos and Martonosi, Margaret R. and Chong, Frederic T.},
  booktitle={2022 IEEE International Symposium on High-Performance Computer Architecture (HPCA)}, 
  title={{SupermarQ: A Scalable Quantum Benchmark Suite}}, 
  year={2022},
  volume={},
  number={},
  pages={587-603},
  doi={10.1109/HPCA53966.2022.00050}
  }

@inproceedings{SC_similarity,
author = {Li, Lingda and Flynn, Thomas and Hoisie, Adolfy},
title = {{Learning Generalizable Program and Architecture Representations for Performance Modeling}},
year = {2024},
isbn = {9798350352917},
publisher = {IEEE Press},
url = {https://doi.org/10.1109/SC41406.2024.00072},
doi = {10.1109/SC41406.2024.00072},
booktitle = {Proceedings of the International Conference for High Performance Computing, Networking, Storage, and Analysis},
articleno = {66},
numpages = {15},
location = {Atlanta, GA, USA},
series = {SC '24}
}

@inproceedings{ICS_Similarity,
author = {Tr\"{u}mper, Lukas and Ben-Nun, Tal and Schaad, Philipp and Calotoiu, Alexandru and Hoefler, Torsten},
title = {{Performance Embeddings: A Similarity-Based Transfer Tuning Approach to Performance Optimization}},
year = {2023},
isbn = {9798400700569},
publisher = {Association for Computing Machinery},
address = {New York, NY, USA},
url = {https://doi.org/10.1145/3577193.3593714},
doi = {10.1145/3577193.3593714},
booktitle = {Proceedings of the 37th ACM International Conference on Supercomputing},
pages = {50–62},
numpages = {13},
location = {Orlando, FL, USA},
series = {ICS '23}
}

@article{qiu2026passbypassoptimizationintentdrivenir,
  title   = {{Beyond Pass-by-Pass Optimization: Intent-Driven IR Optimization with Large Language Models}},
  author  = {Qiu, Lei and Yang, Zi and Lyu, Fang and Zhong, Ming and Cui, Huimin and Feng, Xiaobing},
  journal = {arXiv preprint arXiv:2602.18511},
  year    = {2026},
  doi     = {10.48550/arXiv.2602.18511},
  url     = {https://arxiv.org/abs/2602.18511}
}

@inproceedings{quantum_tuning_cibda,
author = {Liu, Yi and Jin, Yuqiong and Xu, Jinchen},
title = {{A Portable Auto-Tuning Framework for Quantum Compilation Optimization Based on D3QN}},
year = {2025},
isbn = {9798400713163},
publisher = {Association for Computing Machinery},
address = {New York, NY, USA},
url = {https://doi.org/10.1145/3746709.3746950},
doi = {10.1145/3746709.3746950},
booktitle = {Proceedings of the 2025 6th International Conference on Computer Information and Big Data Applications},
pages = {1422–1428},
numpages = {7},
location = {
},
series = {CIBDA '25}
}

@article{tosem_circuits,
author = {Ma, Ning and Li, Heng},
title = {{Understanding and Estimating the Execution Time of Quantum Circuits}},
year = {2025},
publisher = {Association for Computing Machinery},
address = {New York, NY, USA},
issn = {1049-331X},
url = {https://doi.org/10.1145/3778031},
doi = {10.1145/3778031},
note = {Just Accepted},
journal = {ACM Trans. Softw. Eng. Methodol.},
month = nov
}

@article{PhysRevLett_qa,
  title = {{Quantum Algorithm for Linear Systems of Equations}},
  author = {Harrow, Aram W. and Hassidim, Avinatan and Lloyd, Seth},
  journal = {Phys. Rev. Lett.},
  volume = {103},
  issue = {15},
  pages = {150502},
  numpages = {4},
  year = {2009},
  month = {Oct},
  publisher = {American Physical Society},
  doi = {10.1103/PhysRevLett.103.150502},
  url = {https://link.aps.org/doi/10.1103/PhysRevLett.103.150502}
}

@article{openqasm2,
      title={{Open Quantum Assembly Language}}, 
      author={Andrew W. Cross and Lev S. Bishop and John A. Smolin and Jay M. Gambetta},
      year={2017},
      journal = {arXiv preprint arXiv:1707.03429},
      doi     = {10.48550/arXiv.1707.03429},
      url={https://arxiv.org/abs/1707.03429}, 
}

@misc{Web:removebarriers,
  author = {{IBM}},
  title  = {{(2025) RemoveBarriers Pass}},
  note   = {Accessed: Mar 8th 2026. [Online]. Available: \url{https://quantum.cloud.ibm.com/docs/en/api/qiskit/2.2/qiskit.transpiler.passes.RemoveBarriers}}
}

@misc{Web:fakebackends,
  author = {{IBM}},
  title  = {{(2026) Fake Provider}},
  note   = {Accessed: Jan 15th 2026. [Online]. Available: \url{https://quantum.cloud.ibm.com/docs/en/api/qiskit-ibm-runtime/fake-provider}}
}

@article{douze2024faisslibrary,
  title   = {{The Faiss Library}},
  author  = {Douze, Matthijs and Guzhva, Alexandr and Deng, Chengqi and Johnson, Jeff and Szilvasy, Gergely and Mazar{\'e}, Pierre-Emmanuel and Lomeli, Maria and Hosseini, Lucas and J{\'e}gou, Herv{\'e}},
  journal = {arXiv preprint arXiv:2401.08281},
  year    = {2024},
  doi     = {10.48550/arXiv.2401.08281},
  url     = {https://arxiv.org/abs/2401.08281}
}

@article{chen2022veriqbenchbenchmarkmultipletypes,
  title   = {{VeriQBench: A Benchmark for Multiple Types of Quantum Circuits}},
  author  = {Chen, Kean and Fang, Wang and Guan, Ji and Hong, Xin and Huang, Mingyu and Liu, Junyi and Wang, Qisheng and Ying, Mingsheng},
  journal = {arXiv preprint arXiv:2206.10880},
  year    = {2022},
  doi     = {10.48550/arXiv.2206.10880},
  url     = {https://arxiv.org/abs/2206.10880}
}

@INPROCEEDINGS{autophase,
  author={Huang, Qijing and Haj-Ali, Ameer and Moses, William and Xiang, John and Stoica, Ion and Asanovic, Krste and Wawrzynek, John},
  booktitle={2019 IEEE 27th Annual International Symposium on Field-Programmable Custom Computing Machines (FCCM)}, 
  title={{AutoPhase: Compiler Phase-Ordering for HLS with Deep Reinforcement Learning}}, 
  year={2019},
  volume={},
  number={},
  pages={308-308},
  doi={10.1109/FCCM.2019.00049}}

@inproceedings{oopsla_phase,
author = {Kulkarni, Sameer and Cavazos, John},
title = {Mitigating the compiler optimization phase-ordering problem using machine learning},
year = {2012},
isbn = {9781450315616},
publisher = {Association for Computing Machinery},
address = {New York, NY, USA},
url = {https://doi.org/10.1145/2384616.2384628},
doi = {10.1145/2384616.2384628},
booktitle = {Proceedings of the ACM International Conference on Object Oriented Programming Systems Languages and Applications},
pages = {147–162},
numpages = {16},
location = {Tucson, Arizona, USA},
series = {OOPSLA '12}
}

@inproceedings{cc_phase,
author = {Han, Ruobing and Kim, Hyesoon},
title = {{Exponentially Expanding the Phase-Ordering Search Space via Dormant Information}},
year = {2024},
isbn = {9798400705076},
publisher = {Association for Computing Machinery},
address = {New York, NY, USA},
url = {https://doi.org/10.1145/3640537.3641582},
doi = {10.1145/3640537.3641582},
booktitle = {Proceedings of the 33rd ACM SIGPLAN International Conference on Compiler Construction},
pages = {250–261},
numpages = {12},
location = {Edinburgh, United Kingdom},
series = {CC 2024}
}

@INPROCEEDINGS{ipdps_phase,
  author={Zhao, Jiayu and Xia, Chunwei and Wang, Zheng},
  booktitle={2025 IEEE International Parallel and Distributed Processing Symposium (IPDPS)}, 
  title={{Leveraging Compilation Statistics for Compiler Phase Ordering}}, 
  year={2025},
  volume={},
  number={},
  pages={533-545},
  doi={10.1109/IPDPS64566.2025.00054}}

@inproceedings{internetware_phase,
author = {Chen, Yihan and Chen, Huanhuan and Yao, Yuan and Yu, Ping and Xu, Feng and Ma, Xiaoxing},
title = {{Exploiting Booster Pass Chain for Compiler Phase Ordering}},
year = {2025},
isbn = {9798400719264},
publisher = {Association for Computing Machinery},
address = {New York, NY, USA},
url = {https://doi.org/10.1145/3755881.3755899},
doi = {10.1145/3755881.3755899},
booktitle = {Proceedings of the 16th International Conference on Internetware},
pages = {175–185},
numpages = {11},
location = {
},
series = {Internetware '25}
}

@article{QASMBench,
author = {Li, Ang and Stein, Samuel and Krishnamoorthy, Sriram and Ang, James},
title = {{QASMBench: A Low-Level Quantum Benchmark Suite for NISQ Evaluation and Simulation}},
year = {2023},
issue_date = {June 2023},
publisher = {Association for Computing Machinery},
address = {New York, NY, USA},
volume = {4},
number = {2},
url = {https://doi.org/10.1145/3550488},
doi = {10.1145/3550488},
journal = {ACM Transactions on Quantum Computing},
month = feb,
articleno = {10},
numpages = {26}
}

@article{javadiabhari2024quantumcomputingqiskit,
  title   = {{Quantum Computing with Qiskit}},
  author  = {Javadi-Abhari, Ali and Treinish, Matthew and Krsulich, Kevin and Wood, Christopher J. and Lishman, Jake and Gacon, Julien and Martiel, Simon and Nation, Paul D. and Bishop, Lev S. and Cross, Andrew W. and Johnson, Blake R. and Gambetta, Jay M.},
  journal = {arXiv preprint arXiv:2405.08810},
  year    = {2024},
  doi     = {10.48550/arXiv.2405.08810},
  url     = {https://arxiv.org/abs/2405.08810}
}

@misc{Web:qiskit_opt_pass,
	author = {{IBM}},
	title = {{(2025) Optimization Passes in Qiskit Transpiler}},
    note   = {Accessed: Jan 3rd 2026. [Online]. Available: \url{https://quantum.cloud.ibm.com/docs/en/api/qiskit/2.2/transpiler_passes#optimizations}}
}

@inproceedings{SABRE,
author = {Li, Gushu and Ding, Yufei and Xie, Yuan},
title = {{Tackling the Qubit Mapping Problem for NISQ-Era Quantum Devices}},
year = {2019},
isbn = {9781450362405},
publisher = {ACM},
address = {NY, USA},
booktitle = {Proceedings of the Twenty-Fourth International Conference on Architectural Support for Programming Languages and Operating Systems},
pages = {1001–1014},
numpages = {14},
location = {Providence, RI, USA},
series = {ASPLOS '19}
}

@misc{Web:eigen_ai,
  author = {{Eigen AI}},
  title  = {{ (2026) High‑Performance AI for Enterprises}},
  note   = {Accessed: May 14th 2026. [Online]. Available: \url{https://www.eigenai.com/}}
}

@article{Sivarajah_2021,
doi = {10.1088/2058-9565/ab8e92},
url = {https://doi.org/10.1088/2058-9565/ab8e92},
year = {2020},
month = {nov},
publisher = {IOP Publishing},
volume = {6},
number = {1},
pages = {014003},
author = {Sivarajah, Seyon and Dilkes, Silas and Cowtan, Alexander and Simmons, Will and Edgington, Alec and Duncan, Ross},
title = {{tket: a retargetable compiler for NISQ devices}},
journal = {Quantum Science and Technology}
}

@article{bench_qasm,
author = {Nation, Paul and Saki, Abdullah Ash and Brandhofer, Sebastian and Bello, Luciano and Garion, Shelly and Treinish, Matthew and Javadi-Abhari, Ali},
year = {2025},
month = {April},
pages = {427-435},
title = {Benchmarking the performance of quantum computing software for quantum circuit creation, manipulation and compilation},
volume = {5},
journal = {Nature Computational Science},
doi = {10.1038/s43588-025-00792-y}
}

@ARTICLE{quantum_finance,
  author={Chang, Yen-Jui and Sie, Ming-Fong and Liao, Shih-Wei and Chang, Ching-Ray},
  journal={IEEE Nanotechnology Magazine}, 
  title={{The Prospects of Quantum Computing for Quantitative Finance and Beyond}}, 
  year={2023},
  volume={17},
  number={2},
  pages={31-37},
  doi={10.1109/MNANO.2023.3249501}}

@inproceedings{quest,
author = {Wang, Hanrui and Liang, Zhiding and Gu, Jiaqi and Li, Zirui and Ding, Yongshan and Jiang, Weiwen and Shi, Yiyu and Pan, David Z. and Chong, Frederic T. and Han, Song},
title = {{TorchQuantum Case Study for Robust Quantum Circuits}},
year = {2022},
isbn = {9781450392174},
publisher = {Association for Computing Machinery},
address = {New York, NY, USA},
url = {https://doi.org/10.1145/3508352.3561118},
doi = {10.1145/3508352.3561118},
booktitle = {Proceedings of the 41st IEEE/ACM International Conference on Computer-Aided Design},
articleno = {136},
numpages = {9},
location = {San Diego, California},
series = {ICCAD '22}
}

@article{qse_roadmap,
author = {Murillo, Juan Manuel and Garcia-Alonso, Jose and Moguel, Enrique and Barzen, Johanna and Leymann, Frank and Ali, Shaukat and Yue, Tao and Arcaini, Paolo and P\'{e}rez-Castillo, Ricardo and Garc\'{\i}a-Rodr\'{\i}guez de Guzm\'{a}n, Ignacio and Piattini, Mario and Ruiz-Cort\'{e}s, Antonio and Brogi, Antonio and Zhao, Jianjun and Miranskyy, Andriy and Wimmer, Manuel},
title = {{Quantum Software Engineering: Roadmap and Challenges Ahead}},
year = {2025},
issue_date = {June 2025},
publisher = {Association for Computing Machinery},
address = {New York, NY, USA},
volume = {34},
number = {5},
issn = {1049-331X},
url = {https://doi.org/10.1145/3712002},
doi = {10.1145/3712002},
journal = {ACM Transactions on Software Engineering and Methodology},
month = may,
articleno = {154},
numpages = {48}
}

@ARTICLE{qse_test1,
  author={Wang, Xinyi and Ali, Shaukat and Yue, Tao and Arcaini, Paolo},
  journal={IEEE Transactions on Software Engineering}, 
  title={{Quantum Approximate Optimization Algorithm for Test Case Optimization}}, 
  year={2024},
  volume={50},
  number={12},
  pages={3249-3264},
  doi={10.1109/TSE.2024.3479421}
}

@article{qse_test2,
author = {Xia, Shangzhou and Zhao, Jianjun and Zhang, Fuyuan and Guo, Xiaoyu},
title = {{Quantum Concolic Testing}},
year = {2025},
issue_date = {July 2025},
publisher = {Association for Computing Machinery},
address = {New York, NY, USA},
volume = {2},
number = {ISSTA},
url = {https://doi.org/10.1145/3728926},
doi = {10.1145/3728926},
journal = {Proc. ACM Softw. Eng.},
month = jun,
articleno = {ISSTA051},
numpages = {21}
}

@article{qse_test3,
author = {Long, Peixun and Zhao, Jianjun},
title = {{Testing Multi-Subroutine Quantum Programs: From Unit Testing to Integration Testing}},
year = {2024},
issue_date = {July 2024},
publisher = {Association for Computing Machinery},
address = {New York, NY, USA},
volume = {33},
number = {6},
issn = {1049-331X},
url = {https://doi.org/10.1145/3656339},
doi = {10.1145/3656339},
journal = {ACM Trans. Softw. Eng. Methodol.},
month = jun,
articleno = {147},
numpages = {61}
}

@article{qse_repair,
author = {Tan, Siwei and Lu, Liqiang and Xiang, Debin and Chu, Tianyao and Lang, Congliang and Chen, Jintao and Hu, Xing and Yin, Jianwei},
title = {{HornBro: Homotopy-Like Method for Automated Quantum Program Repair}},
year = {2025},
issue_date = {July 2025},
publisher = {Association for Computing Machinery},
address = {New York, NY, USA},
volume = {2},
number = {FSE},
url = {https://doi.org/10.1145/3715751},
doi = {10.1145/3715751},
journal = {Proc. ACM Softw. Eng.},
month = jun,
articleno = {FSE034},
numpages = {23}
}

@inproceedings{tuniq,
author = {Hasanat, Mohammad Abrarul and Ludmir, Jason and Patel, Tirthak and Roy, Rohan Basu},
title = {{TuniQ: Autotuning Compilation Passes for Quantum Workloads at Scale for Effectiveness and Efficiency}},
year = {2026},
isbn = {9798400725227},
publisher = {Association for Computing Machinery},
address = {New York, NY, USA},
url = {https://doi.org/10.1145/3797905.3807862},
doi = {10.1145/3797905.3807862},
booktitle = {Proceedings of the 40th ACM International Conference on Supercomputing},
pages = {1322–1336},
numpages = {15},
location = {
},
series = {ICS '26}
}

@INPROCEEDINGS{ASE_QuanBench,
  author={Guo, Xiaoyu and Wang, Minggu and Zhao, Jianjun},
  booktitle={2025 40th IEEE/ACM International Conference on Automated Software Engineering (ASE)}, 
  title={{QuanBench: Benchmarking Quantum Code Generation with Large Language Models}}, 
  year={2025},
  volume={},
  number={},
  pages={2657-2669},
  doi={10.1109/ASE63991.2025.00218}
}

@INPROCEEDINGS{qse_test4,
  author={Wang, Jiyuan and Zhang, Qian and Xu, Guoqing Harry and Kim, Miryung},
  booktitle={2021 36th IEEE/ACM International Conference on Automated Software Engineering (ASE)}, 
  title={{QDiff: Differential Testing of Quantum Software Stacks}}, 
  year={2021},
  volume={},
  number={},
  publisher = {IEEE Press},
  address = {Melbourne, Australia},
  pages={692-704},
  doi={10.1109/ASE51524.2021.9678792}
}

@inproceedings{qse_codegen2,
author = {Campbell, Charlie and Chen, Hao (Mark) and Luk, Wayne and Fan, Hongxiang},
title = {{Enhancing LLM-Based Quantum Code Generation with Multi-Agent Optimization and Quantum Error Correction}},
year = {2025},
isbn = {9798331503048},
publisher = {IEEE Press},
url = {https://doi.org/10.1109/DAC63849.2025.11133316},
doi = {10.1109/DAC63849.2025.11133316},
booktitle = {Proceedings of the 62nd Annual ACM/IEEE Design Automation Conference},
articleno = {414},
numpages = {7},
location = {San Francisco, California, United States},
series = {DAC '25}
}

@INPROCEEDINGS{ISCA_qasmbench1,
  author={Lin, Wan-Hsuan and Tan, Daniel Bochen and Cong, Jason},
  booktitle={2025 IEEE International Symposium on High Performance Computer Architecture (HPCA)}, 
  title={{Reuse-Aware Compilation for Zoned Quantum Architectures Based on Neutral Atoms}}, 
  year={2025},
  volume={},
  number={},
  pages={127-142},
  doi={10.1109/HPCA61900.2025.00021}
}

@inproceedings{asplos_qasmbench3,
author = {Hua, Fei and Jin, Yuwei and Chen, Yanhao and Vittal, Suhas and Krsulich, Kevin and Bishop, Lev S. and Lapeyre, John and Javadi-Abhari, Ali and Zhang, Eddy Z.},
title = {{CaQR: A Compiler-Assisted Approach for Qubit Reuse through Dynamic Circuit}},
year = {2023},
isbn = {9781450399180},
publisher = {Association for Computing Machinery},
address = {New York, NY, USA},
url = {https://doi.org/10.1145/3582016.3582030},
doi = {10.1145/3582016.3582030},
booktitle = {Proceedings of the 28th ACM International Conference on Architectural Support for Programming Languages and Operating Systems, Volume 3},
pages = {59–71},
numpages = {13},
location = {Vancouver, BC, Canada},
series = {ASPLOS 2023}
}

@inproceedings{PLDI_qasmbench4,
author = {Tao, Runzhou and Shi, Yunong and Yao, Jianan and Li, Xupeng and Javadi-Abhari, Ali and Cross, Andrew W. and Chong, Frederic T. and Gu, Ronghui},
title = {{Giallar: push-button verification for the qiskit Quantum compiler}},
year = {2022},
isbn = {9781450392655},
publisher = {Association for Computing Machinery},
address = {New York, NY, USA},
url = {https://doi.org/10.1145/3519939.3523431},
doi = {10.1145/3519939.3523431},
booktitle = {Proceedings of the 43rd ACM SIGPLAN International Conference on Programming Language Design and Implementation},
pages = {641–656},
numpages = {16},
location = {San Diego, CA, USA},
series = {PLDI 2022}
}

@inproceedings{liang2024combining,
author = {Liang, Zhiding and Song, Zhixin and Cheng, Jinglei and Ren, Hang and Hao, Tianyi and Yang, Rui and Shi, Yiyu and Li, Tongyang},
title = {{Combining Parameterized Pulses and Contextual Subspace for More Practical VQE}},
year = {2024},
isbn = {9798400706011},
publisher = {Association for Computing Machinery},
address = {New York, NY, USA},
url = {https://doi.org/10.1145/3649329.3656245},
doi = {10.1145/3649329.3656245},
booktitle = {Proceedings of the 61st ACM/IEEE Design Automation Conference},
articleno = {122},
numpages = {6},
location = {San Francisco, CA, USA},
series = {DAC '24}
}
